%% file: DAB_J_Resub_v1.tex
\documentclass[12pt, draftclsnofoot, onecolumn]{IEEEtran}
\IEEEoverridecommandlockouts

\usepackage{cite}
\usepackage{amsmath,amssymb,amsfonts}
\usepackage{graphicx}
\usepackage{textcomp}
\usepackage{xcolor}
\usepackage{algorithm}
\usepackage{algorithmic}
\usepackage{balance}
\usepackage{dsfont}
\usepackage{setspace}
\usepackage{tabularx}
\usepackage[caption=false,font=small]{subfig}

\input{vmr-symbols-vecbold}
\input{standard-macros}

\newcommand{\splitatcommas}[1]{%
  \begingroup
  \ifnum\mathcode`,="8000
  \else
    \begingroup\lccode`~=`, \lowercase{\endgroup
      \edef~{\mathchar\the\mathcode`, \penalty0 \noexpand\hspace{0pt plus 1em}}%
    }\mathcode`,="8000
  \fi
  #1%
  \endgroup
}

\begin{document}

\title{Distortion-Aware Linear Precoding\\ for Massive MIMO~Downlink Systems\\ with Nonlinear Power Amplifiers}

\author{\IEEEauthorblockN{Sina Rezaei Aghdam, \textit{Member, IEEE,}~~ Sven Jacobsson,\\ Ulf Gustavsson, ~~ Giuseppe Durisi, \textit{Senior Member, IEEE,}\\ Christoph Studer, \textit{Senior Member, IEEE,}~~ Thomas Eriksson}
	
\thanks{Sina Rezaei Aghdam, Giuseppe Durisi and Thomas Eriksson are with the Department of Electrical Engineering, Chalmers University of Technology, 41296 Gothenburg, Sweden. (e-mails: \{sinar, durisi, thomase\}@chalmers.se).

Sven Jacobsson and Ulf Gustavsson are with Ericsson Research, 41756 Gothenburg,
Sweden.
(e-mails: \{sven.jacobsson, ulf.gustavsson\}@ericsson.com).
	
Christoph Studer is with the Department of Information Technology and Electrical Engineering, ETH Zurich, Z\"urich, Switzerland.
(e-mail: studer@ethz.ch).
		
The work of Sina Rezaei Aghdam and Thomas Eriksson was performed within the strategic innovation program ``Smarter Electronics Systems'', a common venture by VINNOVA, Formas, and Energimyndigheten.
The work of Sven Jacobsson was supported in part by the Swedish Foundation for Strategic Research under grant ID14-0022, and by VINNOVA within the competence center ChaseOn.

This paper was presented in part at the IEEE ICC 2019 Workshop on millimeter-wave communications for 5G and B5G \cite{aghdam}.
	}

}



\maketitle
\vspace{-1em}
\begin{abstract}
We introduce a framework for linear precoder design over a massive multiple-input multiple-output downlink system in the presence of nonlinear power amplifiers (PAs).
By studying the spatial characteristics of the distortion, we demonstrate that conventional linear precoding techniques steer nonlinear distortions towards the users.
We show that, by taking into account PA nonlinearity, one can design linear precoders that reduce, and in single-user scenarios, even completely remove the distortion transmitted in the direction of the users.
This, however, is achieved at the price of a reduced array gain. To address this issue, we present precoder optimization algorithms that simultaneously take into account the effects of array gain, distortion, multiuser interference, and receiver noise.
Specifically, we derive an expression for the achievable sum rate and propose an iterative algorithm that attempts to find the precoding matrix which maximizes this expression.
Moreover, using a model for PA power consumption, we propose an algorithm that attempts to find the precoding matrix that minimizes the consumed power for a given minimum achievable sum rate.
Our numerical results demonstrate that the proposed distortion-aware precoding techniques provide significant improvements in spectral and energy efficiency compared to conventional linear precoders.
\end{abstract}


\section{Introduction}

Massive multiple-input multiple-output (MIMO) technology has already proven its potential for enhancing spectral efficiency in fifth generation~(5G) cellular networks~\cite{bjornson_book}. 
The core idea is to equip base stations with many individually controllable antennas to serve multiple users over the same time-frequency resource via spatial multiplexing. 
To make massive MIMO commercially viable, the base station should be deployed using low-complexity and inexpensive hardware.
Dealing with hardware impairments is therefore one of the main challenges in making massive MIMO practically feasible; see, e.g., \cite{gustavsson14a, bjornson14a, Papazaf17, abdelghany19}.

Low-cost and low-precision hardware can lead to different impairments, such as nonlinear amplification, quantization noise, phase noise, and in-phase/quadrature (I/Q) imbalance. 
How to model these impairments and how to evaluate the resulting performance loss have been the subject of extensive investigations.
As far as modeling is concerned, two approaches are typically followed in the literature.

The first approach deals with the characterization of a single (or predominant) hardware impairment, which is modeled using accurate parametric models. For example, the impact of power amplifier (PA) nonlinearities on the performance of multi-antenna systems has been investigated in \cite{qi12a,blandino17a,mollen18b}. The work in \cite{moghadam18a} characterizes the performance of millimeter-wave multi-antenna systems in terms of spectral and energy efficiency, in the presence of PA nonlinearities and crosstalk.
The effects of other hardware impairments such as phase noise, I/Q imbalance, and quantization have been investigated in, e.g.,~\cite{khanzadi15a, kolomvakis16a, jacobsson17d}.

The second approach is to evaluate the joint effect of multiple hardware impairments; see, e.g., \cite{bjornson14a, zhang15d, Papazafeiropoulos18, studer10b}. 
In these works, the aggregate distortion is simply modeled as additive Gaussian noise, which is often assumed to be uncorrelated across the antenna array. 
While spatially uncorrelated models of distortion can be accurate in some scenarios (see, e.g., \cite{studer10b}), they are typically unsuitable to characterize the massive MIMO downlink with multiuser precoding \cite{larsson18a,anttila19}.
Aggregate hardware impairment models that capture the inherent spatial correlation of the distortion are therefore of great practical importance. 
An example of such an aggregate hardware impairment model has been presented in~\cite{jacobsson18d}.
This model captures the individual distortions caused by nonlinear amplifiers, as well as phase noise and quantization noise.

In scenarios with spatially correlated distortion, knowledge on the covariance matrix of the distortion can be leveraged to enhance the communication performance.
For instance, distortion-aware linear receivers \cite{bjornson19a,aghdam19b} have shown improved performance with respect to conventional linear receivers by (partially) suppressing the distortion at the price of a reduced array gain. 
Distortion-aware channel estimation techniques have also been proposed recently \cite{Demir19}. These techniques are able to reduce the estimation error by exploiting the statistical properties of the hardware impairments.

In the downlink of multiuser MIMO systems, different transmit preprocessing techniques can be used to mitigate the effects of nonlinear distortion.
Per-antenna digital predistortion (DPD) is widely known as an effective technique to mitigate nonlinear distortion at the output of the PAs \cite{sheikhi2021}.
This approach, however, requires a dedicated DPD block per antenna. 
Alternatively, beam-oriented DPD techniques (see, e.g., \cite{brihuega_asil, abdelaziz17}), perform predistortion prior to the baseband precoding block to linearize the signal received at the users.
Despite their promising performance, these solutions often require oversampling and lead to high computational complexity.
In this work, we show that by exploiting the excess spatial degrees of freedom in massive MIMO systems and incorporating the knowledge of distortion correlation and PA power consumption into the design of the linear precoder, one can suppress the nonlinear distortion in the direction of the users and achieve significant gains over conventional linear precoding schemes.
More specifically, the main contributions of this paper are as follows: 
\begin{itemize}
	\item We study the directivity of the nonlinear PAs' distortion in a single-user setting. We model the nonlinear PAs using memoryless polynomials and analytically derive linear precoders that can null the distortion in the direction of the user, but at the cost of reducing the array gain to the user and increasing the radiation in unwanted directions.
	
	\item In view of the fact that the precoders nulling the distortion in the direction of the users are not necessarily favorable solutions in many scenarios, e.g., at low-to-moderate signal-to-noise ratio (SNR), we propose a linear precoder optimization framework that simultaneously takes into account the effect of distortion, the array gain, and the multiuser interference. We utilize Bussgang's theorem \cite{bussgang52a} to decompose the distorted transmit signal into a linear and an uncorrelated distortion term. We then derive a lower bound on the sum rate and propose an iterative procedure, which we refer to as distortion-aware beamforming (DAB). This procedure computes a linear precoder that maximizes the sum rate under an average total power constraint. We show that the DAB precoder is able to realize considerable gains in terms of spectral efficiency compared to conventional distortion-agnostic precoding techniques, such as maximal ratio transmission (MRT) and zero-forcing~(ZF).

	\item We generalize our framework by further considering the consumed power in the PAs. We adopt a simple PA power consumption model and formulate a precoder optimization framework that minimizes the consumed power under a constraint on the minimum required sum rate. We then propose an iterative algorithm to approximately solve this optimization problem. We refer to the output of this iterative algorithm as the energy-efficient DAB (EE-DAB) precoding matrix.

	\item We conduct extensive numerical experiments to evaluate the efficacy of the proposed framework in a variety of scenarios. We quantify the improvements that can be achieved by DAB and EE-DAB in terms of spectral and energy efficiency, respectively, compared to conventional distortion-unaware precoding techniques. Moreover, we study the out-of-band radiation and demonstrate that the proposed precoder optimization framework can be utilized to enforce that the spectral regrowth due to nonlinear PAs stays below given limits.
\end{itemize}

%
%
%
%

\textit{Paper Outline:} The rest of the paper is organized as follows. 
The system model is described in Section \ref{sec:SM}.
In Section \ref{sec:Bussgang_SE}, we utilize Bussgang's theorem to decompose the PA output into a linear term and an uncorrelated distortion term and use this decomposition to derive an expression for the achievable sum rate.
We analyze the directivity of nonlinear distortion under different choices of linear precoders in Section \ref{sec:Spat_Dir_Dist}. 
Our proposed distortion-aware precoder optimization techniques, i.e., DAB and EE-DAB, are detailed in Sections \ref{sec:Dist_Aw} and \ref{sec:En_Eff}, respectively.
Numerical examples that demonstrate the efficacy of the proposed precoders are given in Section \ref{sec:Numerical} and the paper is concluded in Section \ref{sec:Concl}. 

\textit{Notation:}
Lowercase and uppercase boldface letters denote vectors and matrices, respectively. 
The superscripts $(\cdot)^*$, $(\cdot)^T$, and $(\cdot)^H$ denote complex conjugate, transpose, and Hermitian transpose, respectively.
We use $\mathbb{E}[\cdot]$ to denote expectation. We use $\vecnorm{\veca}$ to denote the $\ell_2$-norm of the vector $\veca$.
The $M \times M$ identity matrix is denoted by $\matI_M$ and the $M \times M$ all-zeros matrix is denoted by $\matzero_{M \times M}$. 
We use $\matA \odot \matB$ to denote the Hadamard (entry-wise) product of two equally-sized matrices $\matA$ and $\matB$;
%
%
%
$\text{tr}(\cdot)$ denotes the trace of a matrix.
Furthermore, $\text{diag}(\textbf{a})$ represents a diagonal matrix that contains the elements of the vector $\textbf{a}$ on its diagonal, and
$\text{diag}(\matA)$ is the main diagonal of a square matrix $\matA$.
The element-wise magnitude of a matrix $\matA$ is represented by $|\matA|$.
We use $\angle a$ to denote the angle of the complex number $a$.
The distribution of a circularly symmetric complex Gaussian random vector with covariance matrix $\matC \in \opC^{M \times M}$ is denoted by $\mathcal{CN}(\matzero , \matC)$.
Finally, we use $\mathds{1}_{\setA}(a)$ to denote the indicator function, which is defined as~$\mathds{1}_{\setA}(a) = 1$ for $a \in \setA$ and $\mathds{1}_{\setA}(a) = 0$ for~$a \notin \setA$.

\textit{Reproducible Research:} The MATLAB codes used to generate the results in this paper will be made available upon completion of the review process.


\section{System Model} \label{sec:SM}
\begin{figure}[t]
	\centering
	\includegraphics[width=.8\columnwidth]{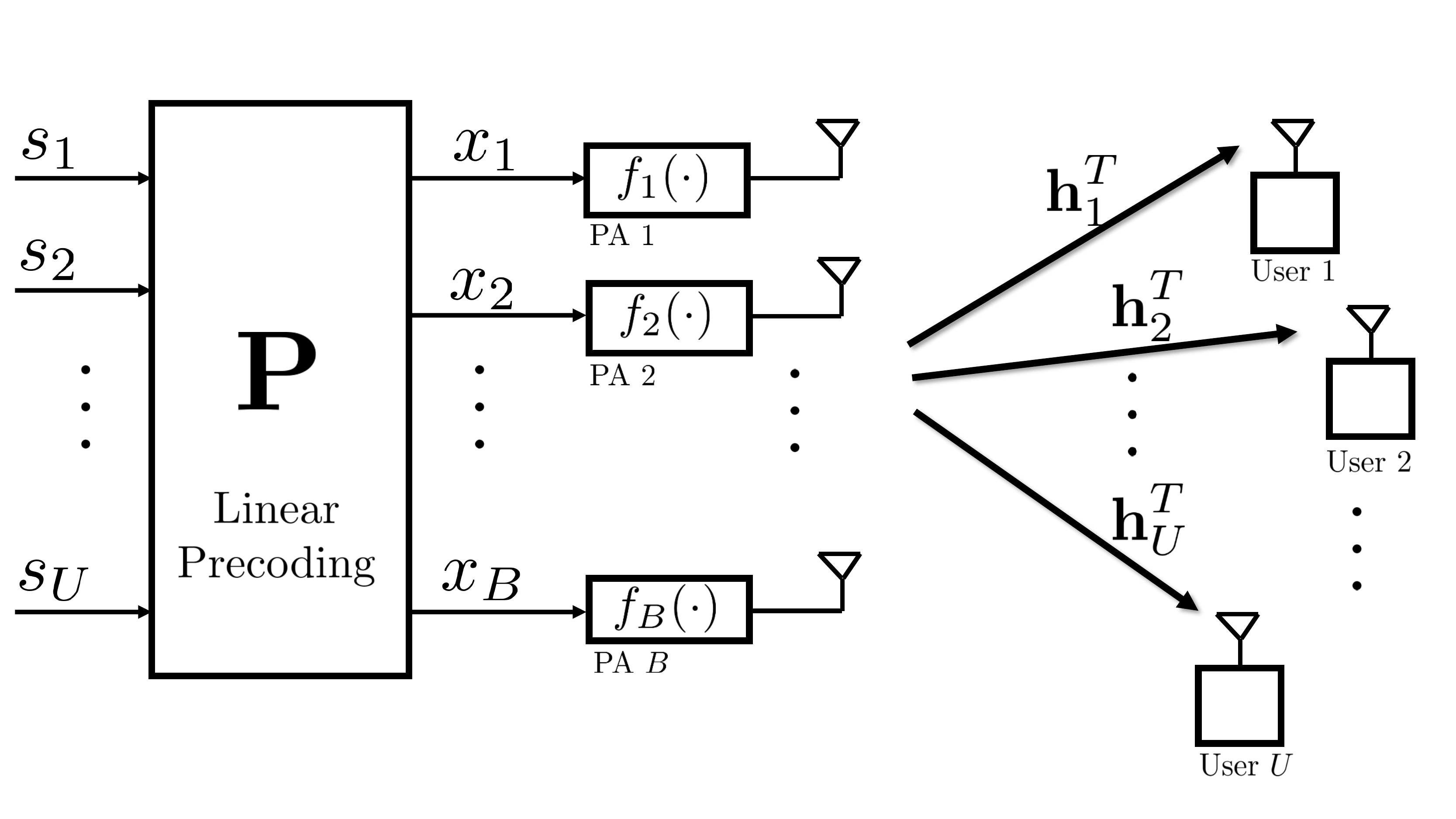}
	\caption{Massive MIMO downlink with linear precoding and nonlinear PAs at the transmitter.}
	\label{fig:system_model}
\end{figure}

We consider a massive MIMO downlink scenario as depicted in Fig. \ref{fig:system_model}. The base station is equipped with $B$ antennas and serves $U$ single-antenna users over the same time-frequency resource.
Each antenna is connected to a nonlinear PA, which introduces distortion to the transmit signal.

\subsection{Channel Input-Output Relationship}
The received signal at the $u$th user is given~by
\begin{equation} \label{eq:Received}
y_u = \vech_u^T f(\vecx) + w_u,
\end{equation}
for $u = 1,\dots,U$, where $\vech_u \in \opC^{B}$ represents the channel between the base station and the $u$th user.
The function $f(\vecx) = [f_1(x_1), \dots, f_B(x_B)]^\text{T}$ describes the output of the PAs. Here, $f_b(\cdot): \opC \rightarrow \opC$ characterizes the memoryless nonlinear behavior of the PA at the $b$th antenna,  $b = 1, \dots, B$.
We consider linear precoding where the precoded signal is given by $\vecx = \matP \vecs$. Here, $\matP =  [\vecp_1, \dots, \vecp_U] \in \mathbb{C}^{B \times U}$ denotes the precoding matrix and $\vecs = [s_1, \dots, s_U]^T \sim \mathcal{CN}(\matzero , \matI_{U})$ models the transmitted symbols. 
Finally, $w_u \sim \mathcal{CN}(0, N_0)$ in \eqref{eq:Received} is additive white Gaussian noise (AWGN).
Throughout the paper, unless stated otherwise, we assume that perfect channel state information (CSI) is known at both the transmitter and the users.\footnote{In Fig. \ref{CDF_with_imp}, we study the impact of imperfect channel state information at the transmitter.}

\subsection{PA Nonlinearity Model}
We model the PAs using a polynomial.~Specifically, the nonlinear behavior of the $b$th PA is modeled by the following $(2K+1)$th order polynomial:\footnote{Since we are mainly interested in the effect of in-band distortion in this paper, even-order terms (which only introduce power components outside the operating frequency band) are neglected in this model. These terms will, however, be reintroduced in a numerical example in Fig. \ref{fig:OOB_emission}, where we study the out-of-band radiation.}
\begin{equation} \label{eq:polyn}
f_b(x_b) = \sum_{k = 0}^{K} \beta_{2k+1}^{(b)} x_b |x_b|^{2k} = \beta_1^{(b)} x_b + \beta_3^{(b)} x_b|x_b|^2 + \dots + \beta_{2K+1}^{(b)} x_b |x_b|^{2K},
\end{equation} 
for $b =  1, \dots, B$, where $\beta_1^{(b)}, \beta_3^{(b)}, \dots, \beta_{2K+1}^{(b)}$ are complex-valued model parameters that capture both amplitude-to-amplitude modulation (AM/AM) and amplitude-to-phase modulation (AM/PM) distortions \cite{schenk2008rf}.
Nonlinear memory effects are typically much weaker than the direct nonlinearities.
Therefore, we have neglected them throughout this work and the analysis of their impact on precoder design is left for future studies.

\subsection{PA Power Consumption Model}
To quantify the power consumption of PAs, we use the model proposed in \cite[Eq. (4)]{Persson13}.
Let us denote the output power of the $b$th PA by $\rho_{\text{tx,}{b}}$ and represent the consumed power at that PA by $\rho_{\text{cons,}{b}}$.
Then the power efficiency can be expressed as
\begin{equation}\label{eq:PAeff}
\eta_b = \frac{\rho_{\text{tx,}{b}}}{\rho_{\text{cons,}{b}}},
\end{equation}
for $b =  1, \dots, B$. 
Through measurements, the authors in \cite{Persson13} have verified that the following equation accurately describes the power efficiency for a variety of PAs:
\begin{equation}\label{eq:PAeff2}
\eta_b = \eta_{\text{max,}{b}} \sqrt{\frac{\rho_{\text{tx,}{b}}}{\rho_{\text{max,}{b}}}}.
\end{equation}
Here, $\eta_{\text{max,}{b}} \in [0, 1]$ is the maximum power efficiency obtained when $\rho_{\text{tx,}{b}} = \rho_{\text{max,}{b}}$. From (\ref{eq:PAeff}) and (\ref{eq:PAeff2}), the consumed power of the $b$th PA can be expressed as
\begin{equation} \label{eq:power_cons}
\rho_{\text{cons,}{b}} = \frac{1}{\eta_{\text{max,}{b}}} \sqrt{\rho_{\text{tx,}{b}}  \rho_{\text{max,}{b}}},
\end{equation}
for $b = 1, \dots, B$ where $\rho_{\text{tx,}{b}} \leq \rho_{\text{max,}{b}}$. In words, the consumed power is proportional to the square root of the output power, which is compatible with results of several other studies; see, e.g., \cite[Eq. (6.93)]{grebennikov2005rf}.

\section{Bussgang Decomposition and Spectral Efficiency} \label{sec:Bussgang_SE}
To analyze the impact of the PA distortion on the performance of the system, we shall use Bussgang's theorem~\cite{bussgang52a} to decompose the nonlinear signal into a scaled linear term and an uncorrelated distortion term. This enables us to derive a lower bound on the sum rate, which we will use as a metric for evaluating the spectral efficiency.

\subsection{Bussgang's Theorem}
By applying Bussgang's theorem~\cite{bussgang52a}, we can, similarly to, e.g.,~\cite{jacobsson18d, bjornson19a, moghadam18a}, rewrite the distorted signal $f(\vecx)$ as
\begin{equation}\label{eq:Bussgang}
f(\vecx) = \matG \vecx + \vece, 
\end{equation}
where $\matG \in \opC^{B \times B}$ is a diagonal matrix containing the Bussgang gains and $\vece \in \opC^B$ is the distortion term which is, by definition, uncorrelated with $\vecx$, i.e., $\mathbb{E}[\vecx\vece^H] = \matzero_{B \times B}$.
The entries of $\matG$ are given by
\begin{equation}\label{eq:Bussgain}
[\matG]_{b,b} = \frac{\Ex{}{f(x_b) x_b^*}}{\Ex{}{|x_b|^2}} = \frac{\sum_{k = 0}^{K} \beta_{2k+1}^{(b)} \Ex{}{ |x_b|^{2k + 2}} }{\Ex{}{|x_b|^2}}.
\end{equation}
%
For the $(2K+1)$th order polynomial model in~\eqref{eq:polyn} and for $\vecx = \matP\vecs$, we obtain $\matG$ in~\eqref{eq:Bussgain} using the moments of the complex Gaussian random variables \cite{reed62a} as 
\begin{equation}
[\matG]_{b,b} = \frac{ \sum_{k = 0}^{K} \beta_{2k+1}^{(b)} (k+1)!~ (\sigma_{x_b}^2)^{k+1}}{\sigma_{x_b}^2} = \sum_{k = 0}^{K} (k+1)!~ \beta_{2k+1}^{(b)}  (\sigma_{x_b}^2)^{k},
\end{equation}
where $\sigma_{x_b}^2$ is the variance of the precoded signal $x_b$ at the $b$th antenna.
Accordingly, the matrix $\matG$ can be expressed as a function of the precoding matrix $\matP$~as
\begin{align}\label{eq:Bussgang_gain} 
\matG(\matP) &= \sum_{k = 0}^{K} (k + 1)!~\matA_{2k+1}  \text{diag}(\matC_{\vecx})^k\\
& = \matA_1 \matI_B + 2 \matA_3 \text{diag}(\matC_{\vecx}) + \dots + (K + 1)!~\matA_{2K+1}  \text{diag}(\matC_{\vecx})^K,
\end{align}
where
\begin{equation}\label{eq:matrixA}
\matA_{2k+1}= \text{diag}\big(\beta_{2k+1}^{(1)}, \dots, \beta_{2k+1}^{(B)}\big),
\end{equation}
for $k = 1, \dots, K$ and $\matC_{\vecx} = \mathbb{E} [\vecx \vecx^H] = \matP\matP^H$ is the covariance matrix of the input. Using (\ref{eq:Bussgang_gain}) and following an approach similar to that used in \cite[Eq.~24]{bjornson19a} and \cite[Eq.~11]{moghadam18a}, the covariance matrix of the distortion $\vece$ can be derived as
\begin{equation}\label{eq:distotion_cov}
\matC_\vece(\matP) = \sum_{k = 1}^{K} \matL_k \matC_{\vecx} \odot |\matC_{\vecx}|^{2k} \matL_k^H,
\end{equation}
where 
\begin{equation}\label{eq:distotion_cov_cont}
\matL_k = \frac{1}{\sqrt{k+1}} \sum_{l = k}^{K}  {l \choose k} (l + 1)!~ \matA_{2l + 1} \text{diag}(\matC_{\vecx})^{l-k}.
\end{equation}
It can be seen from (\ref{eq:distotion_cov}) that if the channel input $\vecx$ has correlated entries, then the distortion will be spatially correlated. In other words, non-zero off-diagonal elements in $\matC_{\vecx}$ (as is generally the case for precoded signal) lead to non-zero off-diagonal elements in the distortion covariance matrix $\matC_\vece$.

\subsection{Achievable Sum Rate}

By inserting (\ref{eq:Bussgang}) into (\ref{eq:Received}), we can write the received signal at the $u$th user as
\begin{IEEEeqnarray}{rCl} 
	y_u 
	&=&  \vech_u^T \matG(\matP)  \vecp_u s_u + \sum_{r \neq u} \vech^T_u \matG(\matP)\vecp_r s_r + \vech_u^T \vece + w_u. \IEEEeqnarraynumspace \label{eq:Received2}
\end{IEEEeqnarray}
Here, $s_u$ is the desired symbol at the $u$th user. Moreover, $\vecp_u$ and $\vecp_r$ denote the $u$th and the $r$th columns of the precoding matrix $\matP$, respectively.
The second, third, and fourth terms on the right-hand side of \eqref{eq:Received2} are the inter-user interference, the received nonlinear distortion, and the AWGN, respectively.
The sum of these three terms can be considered as the effective noise $w_{\text{eff},u} = \sum_{r \neq u} \vech^T_u \matG(\matP)\vecp_r s_r + \vech_u^T \vece + w_u$.
It should be noted that $w_{\text{eff},u}$ is, in general, not Gaussian due to the nonlinearity of the PA, and thus determining the exact sum rate for the input-output model in \eqref{eq:Received2} is not straightforward \cite{lapidoth96b}.
A common approach to evaluate the spectral efficiency is to use the so-called ''auxiliary-channel lower bound'' (see, e.g., \cite[Sec. VI]{arnold06a}) to derive a lower bound on the achievable rate. 
More specifically, in the auxiliary channel, the effective noise $w_{\text{eff},u}$ is replaced by the auxiliary noise $\tilde{w}_{\text{eff},u}$, which is a circularly symmetric complex Gaussian random variable whose variance is the same as that of $w_{\text{eff},u}$.
The capacity of the auxiliary channel can be given in closed form as 
\begin{equation} \label{eq:sum_rate}
R_{\text{sum}}(\matP) = \sum_{u = 1}^{U} \log_2 \left(1 + \text{SINDR}_u(\matP)\right),
\end{equation}
where
\begin{IEEEeqnarray}{rCl}  \label{eq:SINR}
	\text{SINDR}_u(\matP) = \frac{|\vech^T_u \matG(\matP) \vecp_u|^2}{\sum\limits_{r \neq u} |\vech^T_u \matG(\matP) \vecp_r|^2 + \vech^T_u \matC_\vece(\matP) \vech_u^{*} + N_0} \IEEEeqnarraynumspace
\end{IEEEeqnarray}
is the signal-to-interference-noise and distortion ratio (SINDR) at the $u$th user.
This lower bound corresponds to the sum rate achieved with a Gaussian codebook,  assuming that the channel is perfectly known to the users and using a decoder that operates according to the nearest neighbor principle, i.e., that returns the codeword closest to the received signal in the Euclidean norm~\cite{lapidoth96a}.

%

\subsection{Precoding Matrix Normalization}\label{Sec:Normalization}
We utilize the decomposition in \eqref{eq:Bussgang} to introduce a procedure for precoding matrix normalization, which is needed to control the average total transmit power under different choices of the precoding matrix.
In particular, our objective is to find a real-valued normalization factor $\alpha$ such that
\begin{equation}\label{Eq:tot_pow_constraint}
\Ex{}{\vecnorm{f(\alpha \matP \vecs)}^2} = \rho_{\text{tot}},
\end{equation}
where $\rho_{\text{tot}}$ is a positive real value denoting the total power budget.
Let us define $\vecz = f(\alpha \matP \vecs)$ and $\matC_{\vecz} = \mathbb{E}[\vecz \vecz^H]$. Then, the normalization problem can be rewritten as finding $\alpha$ such that $\text{tr}(\matC_{\vecz}) - \rho_{\text{tot}} = 0$, i.e., solving
\begin{equation}\label{eq:normalization}
\text{tr}\left(\matC_{\vece}(\alpha \matP)  + \alpha^2 \matG(\alpha \matP) \matP \matP^H \matG(\alpha \matP)^{H}  \right) - \rho_{\text{tot}}= 0,
\end{equation}
where $\matG(\alpha \matP)$ and $\matC_{\vece}(\alpha \matP)$ are the Bussgang gain and covariance matrix of distortion corresponding to the decomposition of $f(\alpha \matP \vecs)$ as in \eqref{eq:Bussgang}.
Using \eqref{eq:Bussgang_gain}--\eqref{eq:matrixA} and \eqref{eq:distotion_cov}--\eqref{eq:distotion_cov_cont}, we can express~\eqref{eq:normalization} as
\begin{align} \label{eq:normalization_v2} \nonumber
&\sum_{k=1}^{K} \alpha^{2k+4} \text{tr} \bigg(\matL_k \left(\matP \matP^H \odot |\matP \matP^H|^{2k} \right) \matL_k^H \bigg)\\ \nonumber
 &+ \alpha^2 \text{tr} \left( \left(\sum_{k = 0}^{K} \alpha^k (k + 1)!~\matA_{2k+1}  \text{diag}(\matP \matP^H)^k \right) \matP \matP^H \left(\sum_{k = 0}^{K} \alpha^k (k + 1)!~\matA_{2k+1}  \text{diag}(\matP \matP^H)^k \right)^H \right)\\
 &- \rho_{\text{tot}} = 0,
\end{align}
where $\matL_k$ is given in \eqref{eq:distotion_cov_cont}.
Note that the left hand side of \eqref{eq:normalization_v2} is a $(K+2)$th order polynomial in $\xi = \alpha^2$.
Therefore, to solve \eqref{eq:normalization_v2}, all we need to do is to find the roots of this polynomial and then pick the positive real solution for $\xi$, which yields the desired scalar $\alpha = \sqrt{\xi}$.

\section{Spatial Direction of Transmitted Distortion} \label{sec:Spat_Dir_Dist}
In this section, we analyze the directivity of the nonlinear distortion term $\vece$ for different choices of the precoding matrix $\matP$.
To do so, we focus our attention on line-of-sight (LoS) channels. 
Let us assume that the transmit antennas in the system model depicted in Fig. \ref{fig:system_model} are arranged as a uniform linear array (ULA) with $\lambda_c/2$ spacing (where $\lambda_c$ is the carrier wavelength).
Consider a single-path LoS channel between the transmitter and the $u$th user with an angle of departure (AoD) of $\psi_u$. Then, the channel coefficients for the plane-wave model can be expressed as\footnote{In this section, for simplicity, we ignore path-loss effects.}
\begin{equation}\label{eq:LoS_channel}
\left[\vech(\psi_u)\right]_b = e^{-j  \pi(b-1) \cos(\psi_u)},
\end{equation}
for $b = 1,\dots, B$. 

\begin{figure}[t]
	\centering
	\subfloat[MRT precoding; $U=1$ user and $\psi = 100^\circ$.]{\includegraphics[width = .46\textwidth]{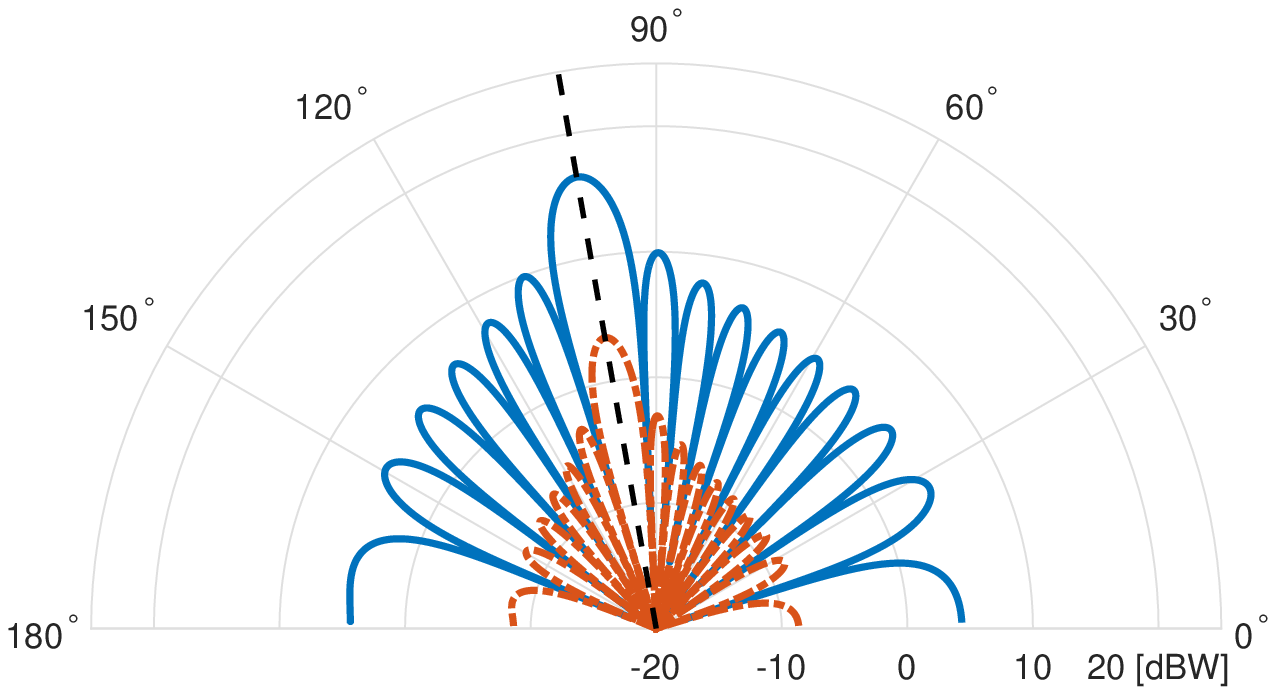}\label{fig:pattern_mrt_1user}} \quad\quad
	\subfloat[MRT precoding; $U=4$ users, $\psi_{1} = 45^\circ$, $\psi_{2} = 100^\circ$, $\psi_{3} = 120^\circ$, and $\psi_{4} = 150^\circ$.]{\includegraphics[width = .46\textwidth]{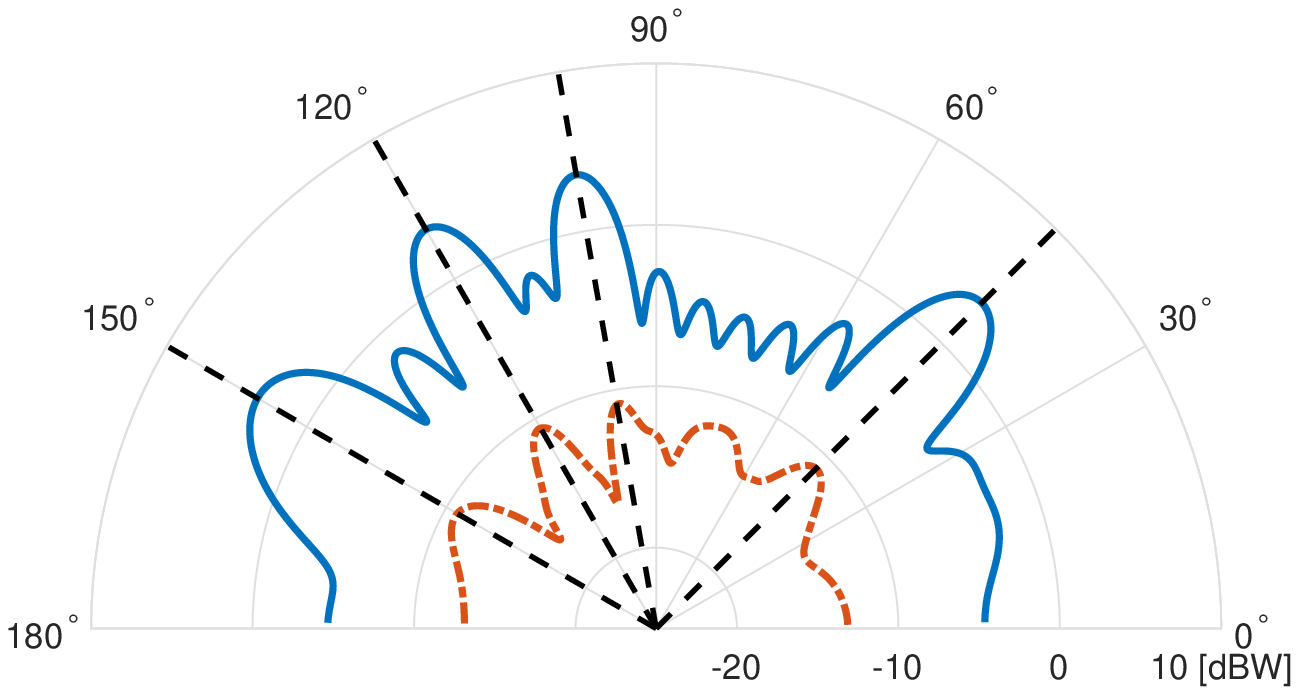}\label{fig:pattern_mrt_4user}} 
	\caption{Far-field radiation pattern for MRT precoding and equal PAs; $B = 16$ antennas, $U \in \{ 1, 4\}$ users. PA coefficients $\beta_1^{(b)} = 0.98$ and $\beta_3^{(b)} = -0.02-0.01j$ for $b = 1, \dots, B$. The precoding matrix is normalized such that the total transmit power is set to $\rho_{\text{tot}} = 43$\,dBm. The blue curves correspond to the radiated linear power $\rho_{\text{lin}}(\psi)$ and the red curves correspond to the distortion power $\rho_{\text{dist}}(\psi)$. MRT precoding steers the distortion in the direction of user(s).}
	\label{fig:MRT_radiation_pattern}
\end{figure}

\subsection{Directivity of Distortion with MRT Precoding}
To study the directivity of the useful signal and distortion, we calculate the power of the linear and distortion terms in different directions $\psi \in [0, 2\pi)$. 
In particular, for a given precoding matrix $\matP$, we evaluate
\begin{equation}\label{eq:dir_lin}
\rho_{\text{lin}}(\psi) = \vech^{T}(\psi) \matG \matP \matP^H \matG^H \vech^{*}(\psi),
\end{equation}
and
\begin{equation}\label{eq:dir_dist}
\rho_{\text{dist}}(\psi) = \vech^{T}(\psi) \matC_\vece \vech^{*}(\psi),
\end{equation}
which correspond to the radiated linear power and the distortion power, respectively. For the sake of simplicity, throughout this section, we consider only a third-order nonlinearity, namely, $K = 1$ in (\ref{eq:polyn}), for which the Bussgang gain in \eqref{eq:Bussgang_gain}--\eqref{eq:matrixA} and the distortion covariance matrix in \eqref{eq:distotion_cov}--\eqref{eq:distotion_cov_cont} reduce to \cite{bjornson19a}
\begin{equation}\label{eq:BG_3rd_order}
\matG = \matA_1 \matI_B + 2\matA_3 \text{diag}(\matP\matP^H),
\end{equation}
and
\begin{equation}\label{eq:Ce_3rd_order}
\matC_\vece(\matP) = 2 \matA_3 \left(\matP \matP^H  \odot |\matP\matP^H|^2 \right) \matA_3^*,
\end{equation}
respectively. 
Let us first consider the special case with $U = 1$ user and equal PAs at different antennas.
In this case, the MRT precoder is given by $\vecp = \alpha \vech^{*}/\|\vech\|$ where $\alpha$ is the normalization factor that has been described in Section \ref{Sec:Normalization}.
Note that, under these assumptions, we get $\vecp \vecp^H  \odot |\vecp\vecp^H|^2 = c \vecp \vecp^H$ where $c$ is a real scalar. 
As a result, by replacing \eqref{eq:BG_3rd_order} and \eqref{eq:Ce_3rd_order} in \eqref{eq:dir_lin} and \eqref{eq:dir_dist}, respectively, we observe that the directivity of the distortion matches exactly that of the linear term. 
This is illustrated in Fig. \ref{fig:pattern_mrt_1user}, where we have plotted the far-field radiation pattern for MRT precoding in a setup with $U = 1$ user and $B = 16$ antennas with equal PAs,  $\beta_1^{(b)} = 0.98$ and $\beta_3^{(b)} = -0.02-0.01j$ for $b = 1, \dots, B$.

Similarly, by evaluating \eqref{eq:dir_lin} and \eqref{eq:dir_dist} for MRT precoding in multiuser scenarios, i.e., $U > 1$, where $\vecp_u = \alpha \vech_u^{*}/\|\vech_u\|$ for $u = 1, \dots, U$, we can see that this precoding technique steers the nonlinear distortion in the direction of the users.
This phenomenon is demonstrated for the case $U = 4$ in Fig. \ref{fig:pattern_mrt_4user}.


The above analysis reveals that in the presence of nonlinear PAs, MRT precoding can lead to a considerable performance degradation due to the beamforming of distortion.\footnote{The same statement is true for ZF precoding.}
Motivated by this, in what follows, we seek precoder design approaches that take explicitly into account the directivity of distortion.
In Section \ref{sec:Zero_dist}, we focus on the case of possibly different PAs at different antennas and demonstrate that for single-user scenarios, properly designed precoders can fully eliminate the distortion in the direction of the user at the price of a reduced array gain. We refer to these precoders as zero-distortion precoders.
Later on, in Section \ref{sec:Dist_Aw}, we study the precoder design in a general scenario with multiple users and arbitrary channel models.

\subsection{Zero-Distortion Precoding in Single-User Scenarios}\label{sec:Zero_dist}
We start by considering a simple scenario with a single user, i.e., $U = 1$ and $B = 2$ transmit antennas. To reduce the complexity further, let us consider a third-order nonlinearity model with  $\beta_1^{(1)} = \beta_1^{(2)} = 1$. In the absence of AWGN, we can write the received signal as
\begin{align} \nonumber
y &= \begin{bmatrix} 
h_1 & h_2
\end{bmatrix} \begin{bmatrix}
p_1 s + \beta_3^{(1)} p_1 s |p_1 s|^2 \\ 
p_2 s + \beta_3^{(2)} p_2 s |p_2 s|^2 
\end{bmatrix} \\ \label{U1_B2}
&= ( h_1 p_1 + h_2 p_2) s + (\beta_3^{(1)} h_1 p_1 |p_1|^2 + \beta_3^{(2)} h_2 p_2 |p_2|^2 ) s|s|^2,
\end{align}
where $p_1$ and $p_2$ are the precoding coefficients associated with the first and the second antenna at the base station, respectively.
It follows from (\ref{U1_B2}) that, to have zero distortion at the UE, the precoding coefficients must satisfy
\begin{equation} \label{zero_dist_U2}
\beta_3^{(1)} h_1 p_1 |p_1|^2 + \beta_3^{(2)} h_2 p_2 |p_2|^2 = 0.
\end{equation}
For the LoS channel given in \eqref{eq:LoS_channel} with a single user at angle $\psi$, the following precoder is a solution of (\ref{zero_dist_U2}):\footnote{This solution can also be found by setting $\rho_{\text{dist}}(\psi)$ to zero.}
\begin{equation}
\vecp = \alpha \begin{bmatrix} \label{zero_dist_precoder}
1 \\ \bigg|\frac{\beta_3^{(1)}}{\beta_3^{(2)}}\bigg|^{\frac{1}{3}} e^{j  \left(\pi \cos(\psi) + \pi + \angle \frac{\beta_3^{(1)}}{\beta_3^{(2)}}\right)}
\end{bmatrix},
\end{equation}
where $\alpha$ is the normalization factor (as described in Section \ref{Sec:Normalization}).
Note that if $\beta_3^{(1)} = \beta_3^{(2)}$, namely, when the PAs at the two antenna ports are identical, the received useful (linear) signal is also zero. Thus, no useful signal is received by the user.
Therefore, in order for this precoder to yield positive achievable rates over a LoS channel, PAs with different distortion profiles are needed at different antenna ports. 

\begin{figure}[t]
	\centering
	\subfloat[MRT; $U=1$ user and $\psi = 100^\circ$.]{\includegraphics[width = .46\textwidth]{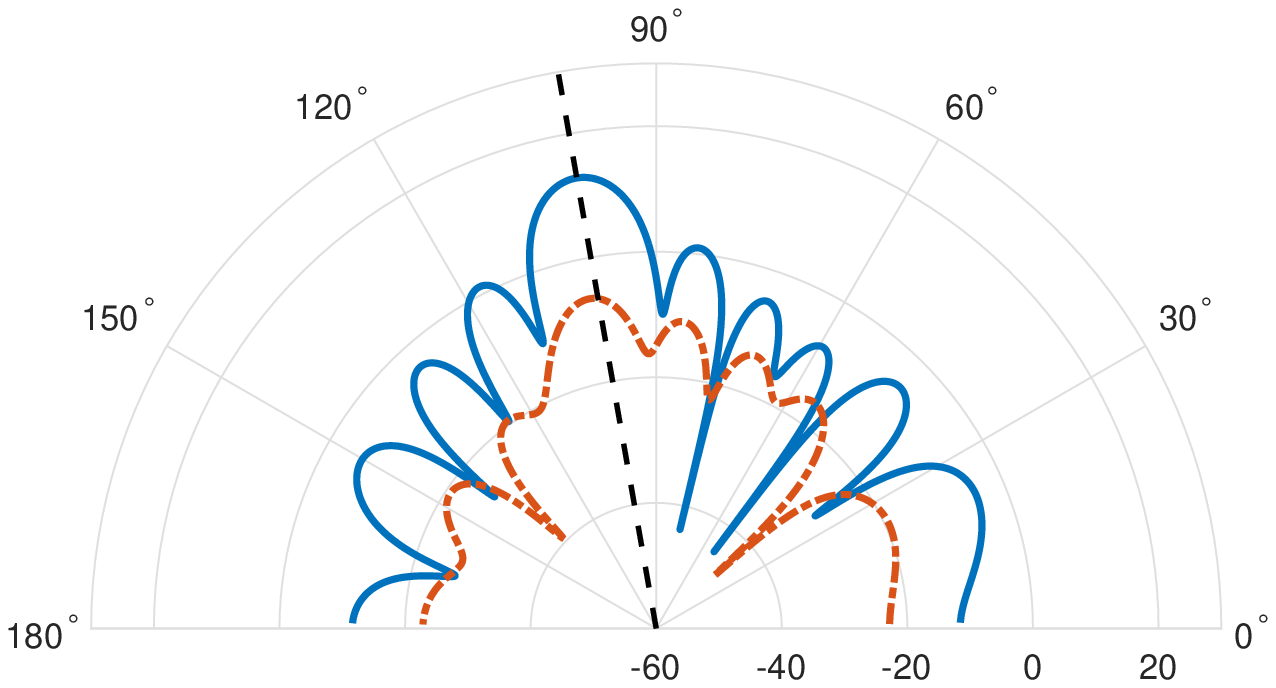}\label{fig:pattern_mrt_dPA}} \quad\quad
	\subfloat[Zero-distortion precoding in \eqref{zerodist_B4}; $U=1$ user and $\psi = 100^\circ$.]{\includegraphics[width = .46\textwidth]{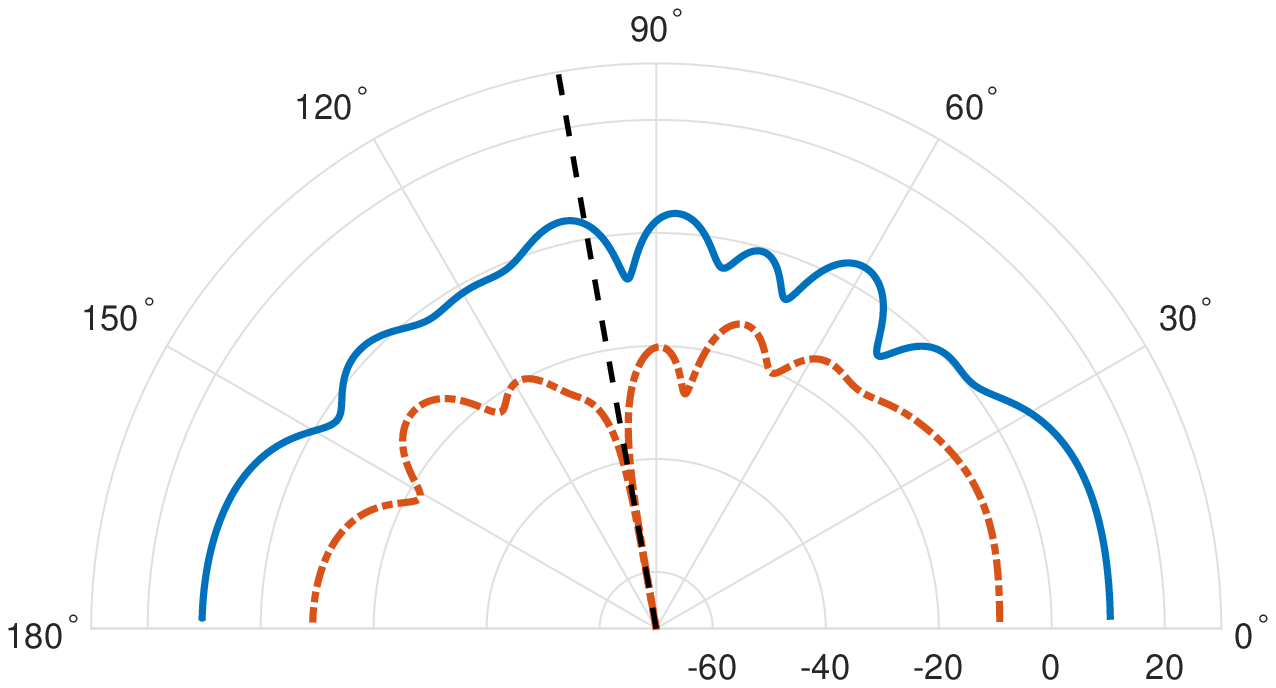}\label{fig:Nodist_radiation_pattern}} 
	\caption{Far-field radiation pattern for MRT and zero-distortion precoding with unequal PAs; $B = 10$ antennas and $U = 1$ user at $\psi = 100$. Total transmit power is set to $\rho_{\text{tot}} = 43$\,dBm. The nonlinearity coefficients are as in Table \ref{tab:non_par}. The blue curve corresponds to the linear term of the transmitted signal and the red curve corresponds to the distortion~term. Zero-distortion precoder nulls the distortion in the direction of the user while reducing array gain by about $25$\,dB compared to MRT.}\label{fig:pattern_dPA}
\end{figure}

For $B > 2$, the problem becomes underdetermined and various precoders may be found that result in zero distortion. For example, when $B$ is even, the following precoder yields zero distortion:

\begin{equation} \label{zerodist_B4}
	\vecp = \alpha \begin{small} \begin{bmatrix}
	1 \\ \bigg|\frac{\beta_3^{(1)}}{\beta_3^{(2)}}\bigg|^{\frac{1}{3}} e^{j  \left(\pi \cos(\psi) + \pi + \angle \frac{\beta_3^{(1)}}{\beta_3^{(2)}}\right) }\\ 1\\ \bigg|\frac{\beta_3^{(3)}}{\beta_3^{(4)}}\bigg|^{\frac{1}{3}} e^{j  \left(\pi \cos(\psi) + \pi + \angle \frac{\beta_3^{(3)}}{\beta_3^{(4)}}\right) }\\ \vdots \\ 1 \\ \bigg|\frac{\beta_3^{(B-1)}}{\beta_3^{(B)}}\bigg|^{\frac{1}{3}} e^{j  \left(\pi \cos(\psi) + \pi + \angle \frac{\beta_3^{(B-1)}}{\beta_3^{(B)}}\right) }
	\end{bmatrix} \end{small}.
\end{equation}

Fig. \ref{fig:pattern_dPA} depicts the far-field radiation pattern for MRT as well as the zero-distortion precoding given in \eqref{zerodist_B4} for a setup with a single user and $B = 10$ antennas with different PAs at different antenna ports.
The nonlinearity parameters for the PAs are given in Table \ref{tab:non_par}.
It can be seen from Fig. \ref{fig:Nodist_radiation_pattern} that the zero-distortion precoder nulls the distortion in the direction of the user at the price of a reduced array gain--- about $25$\,dB reduction compared to that of MRT precoding as depicted in Fig. \ref{fig:pattern_mrt_dPA}.

\begin{table}[b!]
	\caption{PA nonlinearity parameters in the examples demonstrated in Figs \ref{fig:pattern_dPA} and \ref{awesome_plot}.}
	\label{tab:non_par}
	\begin{small}
		\begin{tabular}{ |m{1.2cm}|m{1.2cm}|m{1.2cm}|m{1.2cm}|m{1.2cm}|m{1.2cm}|m{1.2cm}|m{1.2cm}|m{1.2cm}|m{1.2cm}| } 
			\hline
			$\beta_1^{(1)}$ & $\beta_1^{(2)}$ & $\beta_1^{(3)}$ &$\beta_1^{(4)}$& $\beta_1^{(5)}$& $\beta_1^{(6)}$ &$\beta_1^{(7)}$& $\beta_1^{(8)}$ & $\beta_1^{(9)}$&$\beta_1^{(10)}$\\
			\hline
			$1.047$ & $1.026$ & $0.984$ & $1.027$ & $1.003$ & $0.994$ & $1.000$ & $1.011$ & $0.991$ & $1.029$\\
			\hline
			\hline
			$\beta_3^{(1)}$ & $\beta_3^{(2)}$ & $\beta_3^{(3)}$ &$\beta_3^{(4)}$& $\beta_3^{(5)}$& $\beta_3^{(6)}$ &$\beta_3^{(7)}$& $\beta_3^{(8)}$ & $\beta_3^{(9)}$&$\beta_3^{(10)}$\\
			\hline
			$-0.024$ $-0.015j$ & $-0.012$ $-0.035j$ & $-0.043$ $-0.023j$ & $-0.013$ $-0.014j$ & $-0.022$ $-0.028j$ & $-0.038$ $-0.026j$ & $-0.051$ $-0.012j$ & $-0.042$ $-0.021j$ & $-0.030$ $-0.019j$ & $-0.014$ $-0.038j$\\
			\hline
		\end{tabular}
	\end{small}
	
\end{table}

\begin{figure}[t!] 
	\centering
	\noindent
	\includegraphics[width=10cm]{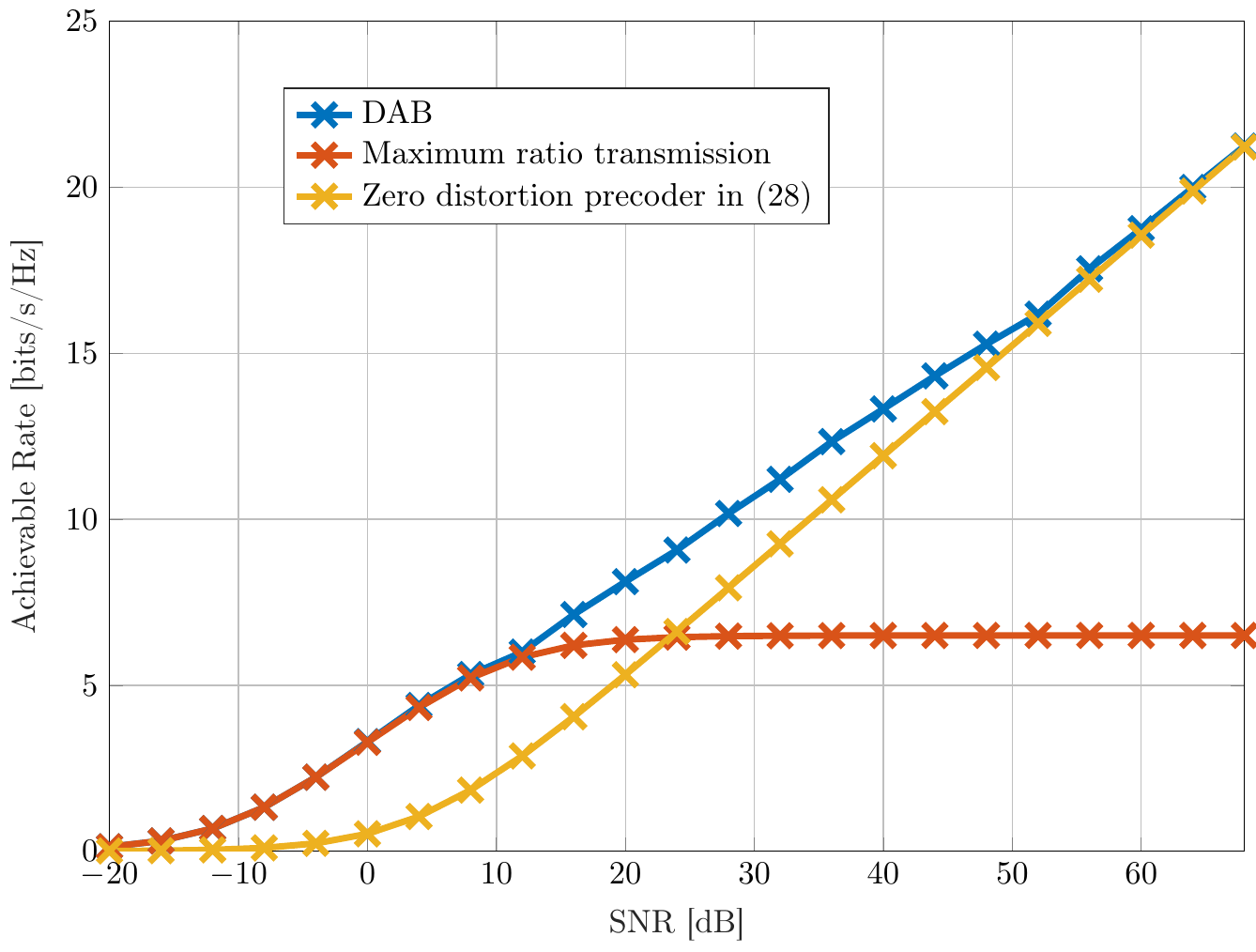}
	\caption{Achievable rates averaged over $500$ realizations of LoS channels in \eqref{eq:LoS_channel} with uniformly randomly distributed AoDs $\psi \in [0, 180)$, $B=10$ antennas at base station and $U=1$ user. The nonlinearity coefficients are as in Table \ref{tab:non_par}. With distortion-aware precoding, the achievable rate becomes a strictly increasing function of the SNR.}
	\label{awesome_plot}
\end{figure}

In Fig. \ref{awesome_plot}, for different SNR values, where we define $\text{SNR}={\rho_{\text{tot}}}/{N_0}$, we evaluate the sum rate
in \eqref{eq:sum_rate} for the precoder in (\ref{zerodist_B4}) as well as the MRT precoder, over LoS channels with $B=10$ and $U = 1$.
It can be seen that the MRT precoder can achieve a higher rate than the zero-distortion precoder at low SNR values where AWGN dominates the nonlinear distortion. However, at higher SNRs, where the nonlinear distortion is dominant, the achievable rate for MRT saturates, whereas the achievable rate for the precoder in (\ref{zerodist_B4}) is a strictly increasing function of SNR.

As a "sneak peek" of what we are going to describe in the next section, we have also included the results achieved by DAB precoding scheme, which achieves an enhanced performance over the whole range of SNR and converges to the zero-distortion precoder at high SNR values.
The algorithm explained in the next section is based on a general precoder design framework that is able to simultaneously take into account the effect of the linear precoder on the array gain, distortion, and multiuser interference.    

\section{Distortion-Aware Precoding} \label{sec:Dist_Aw}
The choice of the precoding matrix $\matP$ affects the sum rate as it controls the array gain, the multiuser interference, and the distortion radiated in the direction of the users (as we saw in Section \ref{sec:Spat_Dir_Dist}).
Motivated by this observation, we now present a framework for optimizing the precoding matrix $\matP$ that uses knowledge of the nonlinearity parameters of PA in addition to CSI and aims to maximize the lower bound on the sum rate in \eqref{eq:sum_rate}.
In particular, we formulate this problem as
\begin{align} \label{eq:MaxRsum}
\begin{array}{lll}
\underset{\matP\in \opC^{B \times U}}{\text{maximize}} & R_{\text{sum}}(\matP)\\
\text{subject to} & \Ex{}{\vecnorm{f(\matP\vecs)}^2} = \rho_{\text{tot}}.
\end{array}
\end{align}
Here, $\rho_{\text{tot}}$ denotes the average total transmit power. 
The power constraint in $\Ex{}{\vecnorm{f(\matP\vecs)}^2} = \rho_{\text{tot}}$ corresponds to a scenario with fixed total transmit power.\footnote{Later on in Fig. \ref{avsr}, we demonstrate how this algorithm can be employed along with a power control strategy.}
Note that~\eqref{eq:MaxRsum} is a non-convex optimization problem since $R_{\text{sum}}(\matP)$ is a non-convex function of $\matP$.
Inspired by the solution in \cite[Sec. V]{palomar05gradient}, in what follows, we solve this problem approximately using an iterative projected gradient ascent algorithm. We refer to this algorithm as DAB. In particular, DAB updates the precoding matrix by taking steps along the steepest ascent direction of the objective function $R_{\text{sum}}(\matP)$. The resulting precoding matrix is then normalized to ensure the feasibility of the solution. 
This procedure can be formulated as \cite[Eq. (72)]{palomar05gradient}
\begin{IEEEeqnarray}{rCl} \label{eq:update_precoding_matrix}
	\widetilde\matP &=& \left[ \matP^{(i-1)} + \mu^{(i-1)} \nabla_\matP R_{\text{sum}}\big(\matP^{(i-1)}\big) \right]_{\opE{\vecnorm{f\left(\matP^{(i-1)}\vecs\right)}^2} = \rho_{\text{tot}}}^{+}\!. \IEEEeqnarraynumspace
\end{IEEEeqnarray}
Here, $i = 1,\dots, I$ is the iteration index, $I$ is the maximum number of iterations, $\mu^{(i)}$ is the step size of the $i$th iteration, and $[\cdot]_{\opE{\vecnorm{f\left(\matP^{(i-1)}\vecs\right)}^2} = \rho_{\text{tot}}}^{+}$ denotes the normalization of the updated precoding matrix, such that the power constraint $\Ex{}{\vecnorm{f(\matP^{(i-1)}\vecs)}^2} = \rho_{\text{tot}}$ is satisfied. The details of this normalization have been explained in Section \ref{Sec:Normalization}.
If $R_\text{sum}(\widetilde\matP) > R_\text{sum}(\matP^{(i-1)})$, we update the precoding matrix to $\matP^{(i)} = \widetilde\matP$ and reset the step size $\mu^{(i)} = \mu^{(0)}$. Otherwise, we do not update the precoding matrix, i.e., $\matP^{(i)} = \matP^{(i-1)}$, and decrease the step size $\mu^{(i)} = \frac{1}{2}\mu^{(i-1)}$. Finally, we choose $\matP_\text{DAB} = \matP^{(I)}$ as the output of the algorithm. We refer to this solution as the DAB precoding matrix.
In Algorithm~1, we summarize the steps required for computing the DAB precoding matrix using a projected gradient~ascent approach.

The computational complexity of Algorithm~1 is dominated by Step~3 and in particular the gradient calculation.
We now explain the procedure for gradient calculation in \eqref{eq:update_precoding_matrix}.
The gradient calculation can be carried out either numerically or through evaluation of the closed-form expression.
In order to compute $ \nabla_\matP R_{\text{sum}}(\matP)$ numerically, we shall evaluate
\begin{equation}\label{eq:grd_numerical}
\big[ \nabla_\matP R_{\text{sum}}(\matP) \big]_{b,u} \approx \frac{\Big(R_{\text{sum}}\big(\matP + \boldsymbol{\Delta}_{\text{r}}^{(b,u)}\big) - R_{\text{sum}}\big(\matP\big)\Big) + j \Big(R_{\text{sum}}\big(\matP + \boldsymbol{\Delta}_{\text{i}}^{(b,u)}\big) - R_{\text{sum}}\big(\matP\big)\Big)}{\delta},
\end{equation}
for $b = 1, \dots, B$ and $u = 1, \dots, U$, where $\delta$ is a small positive constant and $\boldsymbol{\Delta}_{\text{r}}^{(b,u)}$ is a $B \times U$ matrix with $\delta$ on the $b$th row and the $u$th column and zero entries on all other positions.
Similarly, $\boldsymbol{\Delta}_{\text{i}}^{(b,u)}$ is a $B \times U$ matrix whose only non-zero entry is $j \delta$ on the $b$th row and the $u$th column.
A more accurate way of calculating this gradient is via deriving and evaluating closed-form expressions. This case is discussed for the special case of third-order nonlinearity ($K = 1$) in Appendix \ref{sec:App_gradient}.

\begin{algorithm}[t!]
	\begin{spacing}{1.2}
		\caption{Algorithm for computing the distortion-aware beamforming (DAB) precoding matrix.}
		\begin{algorithmic}[1]
			\renewcommand{\algorithmicrequire}{\textbf{Inputs:}}
			\renewcommand{\algorithmicensure}{\textbf{Output:}}
			\REQUIRE $\vech_1, \dots, \vech_U$, $\beta_1^{(b)}, \dots, \beta_K^{(b)}$ for $b=1, \dots, B$, $\rho_\text{tot}$, and $N_0$ 
			\ENSURE  $\matP_\text{DAB}$
			\\ \textit{Initialization}: $\mu^{(0)}$ and $\matP^{(0)}$
			\STATE $R_\text{sum}^{(0)} \leftarrow R_{\text{sum}}(\matP^{(0)})$
			\FOR{$i = 1,\dots,I$}
			\STATE $ \widetilde\matP \leftarrow \left[ \matP^{(i-1)} \!+\! \mu^{(i-1)} \nabla_\matP R_{\text{sum}}\big(\matP^{(i-1)}\big) \right]_{\opE{\vecnorm{f\left(\matP^{(i-1)}\vecs\right)}^2} = \rho_{\text{tot}}}^{+}$
			\STATE $\widetilde{R}_\text{sum} \leftarrow R_{\text{sum}}(\widetilde\matP)$
			\IF {$\widetilde{R}_\text{sum} > R_\text{sum}^{(i-1)}$}
			\STATE $\matP^{(i)} \leftarrow \widetilde\matP$, $R_\text{sum}^{(i)} \leftarrow \widetilde{R}_\text{sum}$, and $\mu^{(i)} \leftarrow \mu^{(0)}$
			\ELSE
			\STATE $\matP^{(i)} \leftarrow \matP^{(i-1)}$, $R_\text{sum}^{(i)} \leftarrow R_\text{sum}^{(i-1)}$, and $\mu^{(i)} \leftarrow \frac{1}{2}\mu^{(i-1)}$
			\ENDIF
			\ENDFOR
			\STATE $\matP_\text{DAB} \leftarrow \matP^{(I)}$
		\end{algorithmic} 
	\end{spacing}
\end{algorithm}

Due to the non-convexity of the problem, Algorithm~1 does not provide global optimality guarantees, i.e., it can return a local maximum. To increase the likelihood of Algorithm~1 to converge to the global maximum, we repeat it with multiple initializations and pick the solution that achieves the highest sum~rate.
Including the conventional precoding matrices, i.e., MRT and ZF, among the set of initializations can help accelerate the convergence, and it guarantees that the DAB precoder does not perform worse than these precoders.
It is worth noting that Algorithm~1 can be optimized in several ways to improve convergence\footnote{For example, by choosing the step sizes $\mu^{(i)}$ according to Armijo's step size rule, improved convergence properties can be achieved \cite{bertsekas1997nonlinear}.} or reduce computational complexity.
However, since the main goal of this paper is to highlight the performance gain that can be achieved by distortion-aware linear precoding, rather than proposing an optimal transmission scheme, we shall not discuss such optimizations.

\section{Energy-Efficient Distortion-Aware Precoding} \label{sec:En_Eff}
In Section \ref{sec:Dist_Aw}, we studied the precoder design with the objective of maximizing the sum rate under a given power constraint.
The average total power constraint in \eqref{eq:MaxRsum} does not take into account the maximum output power constraint of the individual PAs.
In this section, we consider a precoding matrix optimization problem, in which the precoding matrix is subjected to a per-antenna maximum radiated power constraint.
Furthermore, we adopt the model in \eqref{eq:power_cons} to quantify the total base station side consumed power.
This model takes into account the dissipated power in addition to the sum radiated power, which allows for the evaluation and optimization of energy efficiency.
%
%

%
%
A common way for quantifying energy efficiency is via the evaluation of the ratio between throughput and consumed power \cite{feng13survey}
\begin{equation}\label{EE}
\eta_{\text{EE}} = \frac{W R_{\text{sum}}(\matP)}{\rho_{\text{cons,tot}}(\matP)},
\end{equation}
measured in bits per joule. Here, $W$ denotes the bandwidth and $\rho_{\text{cons,tot}}(\matP) = \sum_{b = 1}^{B}  \rho_{\text{cons,}{b}}(\matP)$ is the total consumed power which can be evaluated using \eqref{eq:power_cons}.
In an attempt to maximize energy efficiency, we seek the precoding matrix $\matP$ that minimizes the consumed power while guaranteeing a given minimum required sum rate, i.e.,
\begin{equation}\label{min_sum_rate}
R_{\text{sum}}(\matP) \geq R_0, 
\end{equation}
and under a per-antenna power constraint given by
\begin{equation}\label{max_per_antenna}
\rho_{\text{tx,}{b}} = \Ex{}{|f_{b}(x_{b})|^2} \leq \rho_{\text{max,}{b}},
\end{equation}
for $b = 1, \dots,B$. In \eqref{max_per_antenna}, $x_b$ is the precoded signal input to the $b$th PA and can be obtained by multiplying the $b$th row of the precoding matrix $\matP$ by the transmitted symbols $\vecs$.
The energy efficiency maximization problem can be formulated as
\begin{align} \label{EE_optimization}
\begin{array}{lll}
\underset{\matP\in \opC^{B \times U}}{\text{minimize}} & \rho_{\text{cons,tot}}(\matP)\\
\text{subject to} & \eqref{min_sum_rate}~\text{and}~ \eqref{max_per_antenna}.
\end{array}
\end{align}
%

We solve the nonconvex problem in \eqref{EE_optimization} approximately in two steps.
First, we find the precoding matrix that maximizes the sum rate under a per-antenna power constraint, namely, we maximize $R_{\text{sum}}(\matP)$ subject to \eqref{max_per_antenna}. We refer to this problem as \textit{\textbf{S1}}.
Then, starting from the solution of \textit{\textbf{S1}}, we minimize the consumed power by updating the precoding matrix along the descent direction of the consumed power $\rho_{\text{cons,tot}}(\matP)$ until \eqref{min_sum_rate} is violated or a maximum number of iterations is reached. 
This problem is referred to as \textit{\textbf{S2}} and its output, which is an approximate solution of \eqref{EE_optimization}, is called energy-efficient DAB (EE-DAB) precoding matrix.

With a slight modification, Algorithm~1 in Section \ref{sec:Dist_Aw} can be used to solve \textit{\textbf{S1}}. In particular, we replace \eqref{eq:update_precoding_matrix} (Step 3 in Algorithm~1) with
\begin{IEEEeqnarray}{rCl} \label{eq:update_precoding_matrix_ineq}
	\widetilde\matP &=& \left[ \matP^{(i-1)} + \mu^{(i-1)} \nabla_\matP R_{\text{sum}}\big(\matP^{(i-1)}\big) \right]_{\Ex{}{|f_{b}(x_{b})|^2} \leq \rho_{\text{max,}{b}}}^{+}\!, \IEEEeqnarraynumspace
\end{IEEEeqnarray}
where $[\cdot]_{\Ex{}{|f_{b}(x_{b})|^2} \leq \rho_{\text{max,}{b}}}^{+}$ denotes \textit{occasional} normalization of the $b$th row of the precoding matrix $\matP$ such that \eqref{max_per_antenna} is satisfied.
Specifically, after updating the precoding matrix (line 3 of Algorithm~1), normalization is carried out only if the per-antenna power constraint is violated at one or more antennas. 
For these antennas, the normalization problem is equivalent to finding the values of the scalars $\alpha_{\hat{b}}$ such that $\Ex{}{|f(\alpha_{\hat{b}} x_{\hat{b}} )|^2} = \rho_{\text{max},\hat{b}}$ where $\hat{b}$ denotes the index of an antenna for which \eqref{max_per_antenna} is violated. 
The normalization problem can be solved using the same procedure described in \eqref{eq:normalization}--\eqref{eq:normalization_v2}.
Other than the normalization, all other steps are implemented in the same way as explained in Section \ref{sec:Dist_Aw}.

\begin{algorithm}[t!]
	\begin{spacing}{1.2}
		\caption{Algorithm for computing the energy-efficient DAB (EE-DAB) precoding matrix.}
		\begin{algorithmic}[1]
			\renewcommand{\algorithmicrequire}{\textbf{Inputs:}}
			\renewcommand{\algorithmicensure}{\textbf{Output:}}
			\REQUIRE $R_0$, $\vech_1, \dots, \vech_U$, $\beta_1^{(b)}, \dots, \beta_K^{(b)}$, $\rho_{\text{max,}{b}}$, $\eta_{\text{max,}{b}}$ for $b=1, \dots, B$, and $N_0$
			\ENSURE  $\matP_\text{EE-DAB}$
			\\ \textit{Initialization}: $\mu^{(0)}$ and $\matP^{(\text{S1})}$ from Algorithm~1 (with per-antenna normalization)
			\STATE $\rho_{\text{cons,tot}}^{(0)} \leftarrow \rho_{\text{cons,tot}}(\matP^{(\text{S1})})$ and $i = 1$
			\WHILE{$R_{\text{sum}}(\matP^{(i)}) \geq R_0$ ~ and ~ $i \leq I$}
			\STATE $ \widetilde\matP \leftarrow \left[ \matP^{(i-1)} \!-\! \mu^{(i-1)} \nabla_\matP \rho_{\text{cons,tot}}\big(\matP^{(i-1)}\big) \right]_{\opE{\vecnorm{f\left(x_b\right)}^2} \leq \rho_{\text{max,}{b}}}^{+}$
			\STATE $\widetilde{\rho}_{\text{cons,tot}} \leftarrow \rho_{\text{cons,tot}}(\widetilde\matP)$
			\IF {$\widetilde{\rho}_{\text{cons,tot}} < \rho_{\text{cons,tot}}^{(i-1)}$}
			\STATE $\matP^{(i)} \leftarrow \widetilde\matP$, $\rho_{\text{cons,tot}}^{(i)} \leftarrow \widetilde{\rho}_{\text{cons,tot}}$, and $\mu^{(i)} \leftarrow \mu^{(0)}$
			\ELSE
			\STATE $\matP^{(i)} \leftarrow \matP^{(i-1)}$, $\rho_{\text{cons,tot}}^{(i)} \leftarrow \rho_{\text{cons,tot}}^{(i-1)}$, and $\mu^{(i)} \leftarrow \frac{1}{2}\mu^{(i-1)}$
			\ENDIF
			\STATE $ i = i + 1$
			\ENDWHILE
			\STATE $\matP_\text{DAB} \leftarrow \matP^{(\hat{i})}$ where $\hat{i} = i - 1$ is the final iteration where the stopping criterion is met.
		\end{algorithmic} 
	\end{spacing}
\end{algorithm}

Once \textit{\textbf{S1}} is solved, its solution $\matP^{(\text{S1})}$ is used to initialize the algorithm for solving \textit{\textbf{S2}}. This algorithm seeks the precoding matrix that minimizes the consumed power for a given sum rate no smaller than $R_0$.
Specifically, it updates the precoding matrix using the projected gradient descent, namely, in each iteration,
\begin{IEEEeqnarray}{rCl} \label{eq:update_precoding_matrix_s2}
	\widetilde\matP &=& \left[ \matP^{(i-1)} - \mu^{(i-1)} \nabla_\matP \rho_{\text{cons,tot}}\big(\matP^{(i-1)}\big) \right]_{\Ex{}{|f_{b}(x_{b})|^2} \leq \rho_{\text{max,}{b}}}^{+}\!, \IEEEeqnarraynumspace
\end{IEEEeqnarray}
is calculated for $i = 1,\dots, I$, where $\matP^{(0)} = \matP^{(\text{S1})}$. The gradient $\nabla_\matP \rho_{\text{cons,tot}}\big(\matP^{(i-1)}\big)$ can be evaluated similarly to \eqref{eq:grd_numerical} as
\begin{align}\label{eq:grd_numerical_rhotot} \nonumber
\big[ \nabla_\matP & \rho_{\text{cons,tot}}(\matP) \big]_{b,u} \\& \approx \frac{\Big(\rho_{\text{cons,tot}}\big(\matP + \boldsymbol{\Delta}_{\text{r}}^{(b,u)}\big) - \rho_{\text{cons,tot}}\big(\matP\big)\Big) + j \Big(\rho_{\text{cons,tot}}\big(\matP + \boldsymbol{\Delta}_{\text{i}}^{(b,u)}\big) - \rho_{\text{cons,tot}}\big(\matP\big)\Big)}{\delta}
\end{align}
for $b = 1, \dots, B$ and $u = 1, \dots, U$.
The normalization in \eqref{eq:update_precoding_matrix_s2}, similarly to that of \eqref{eq:update_precoding_matrix_ineq}, is carried out occasionally and on the rows of the precoding matrix that violate \eqref{max_per_antenna}. 
At each iteration, if $\rho_{\text{cons,tot}}(\widetilde\matP) < \rho_{\text{cons,tot}}(\matP^{(i-1)})$, the precoding matrix is updated to $\matP^{(i)} = \widetilde\matP$ and the step size is reset to $\mu^{(i)} = \mu^{(0)}$. Otherwise, the step size is decreased according to $\mu^{(i)} = \frac{1}{2}\mu^{(i-1)}$. This procedure is repeated until either the sum rate goes below $R_0$ or the maximum number of iteration $I$, is reached. This procedure is summarized in Algorithm~2.
%


\section{Numerical Results} \label{sec:Numerical}

We verify the efficacy of the proposed framework by means of numerical simulations.
In our simulations, we focus on millimeter-wave communication scenarios without loss of generality.
We first describe the channel model we use for our simulations (except for the example in Section \ref{OOB}).
Then, we compare the spectral and energy efficiencies achieved by DAB and EE-DAB with those achieved by MRT and ZF precoding.
Finally, we evaluate the out-of-band radiation associated with the proposed precoding techniques.

\subsection{Geometric Channel Model} \label{sec:geo}
In what follows, unless stated otherwise, we adopt a channel model that captures the sparse scattering characteristics of millimeter-wave channels in non-LoS (nLoS) environments, namely, when there is no dominant path.
In this model, which is typically referred to as the geometric channel model, each scatterer contributes to a single path and the channel coefficients can be expressed as\cite{alkhateeb15a}
\begin{equation} \label{eq:channel}
\vech_u = \sqrt{\frac{B}{N_{\text{path}}}} \sum_{\ell = 1}^{N_{\text{path}}} \zeta_{u,\ell} \veca(\psi_{u,\ell}),
\end{equation}
for $u = 1,\dots, U$, where  $N_{\text{path}}$ represents the number of paths/scatterers. 
Here, $\psi_{u,\ell}$ is the AoD for the $\ell$th path, and $\veca(\psi_{u,\ell})$ is the corresponding array response vector. For a ULA with $\lambda_c/2$ antenna spacing, the  $b$th entry of $\veca(\psi_{u,\ell})$ is given by
\begin{equation}
\left[\veca(\psi_{u,\ell})\right]_b = \frac{1}{\sqrt{B}}e^{-j  \pi(b-1) \cos(\psi_{u,\ell})}
\end{equation}
for $b = 1,\dots,B$. 
Furthermore, $\zeta_{u,\ell} \sim \mathcal{CN}(0, \gamma_u^2)$ is the independent and identically distributed (i.i.d.) channel gain (including path loss) corresponding to the $\ell$th path.

Throughout our simulations, we set $N_{\text{path}} = 4$ and assume that the AoD $\psi_{u,\ell}$ is uniformly distributed over the interval $[0^\circ, 180^\circ)$.
We adopt the nLoS path-loss model presented in \cite[Tbl.~I]{akdeniz14a} and, assuming that the system operates at a carrier frequency $f_c = 28$\,GHz (such that $\lambda_c \approx 10.7$\,mm), the path-loss for the user $u$ (at a distance of $d_u$ meters) is calculated using
\begin{equation}
\gamma_u^2 = -72 - 29.2 \log_{10}(d_u) ~~~~~ [\text{dB}].
\end{equation}
We further assume that the users are uniformly distributed in a disk-shaped area with the base station at its center. The minimum and maximum distances from the base station are set to $d_{\text{min}} = 5$ and $d_{\text{max}} = 40$ meters, respectively.
For this setting, the average path loss is about $\gamma_{\text{avg}}^2 = -110$\,dB, which corresponds to a user at the distance of $19.8$ meters from the base station.
Finally, we use the following definition for the average SNR:
\begin{equation}
\text{SNR}_{\text{avg}} = \gamma_{\text{avg}}^2  \frac{\rho_{\text{tot}} }{N_0}.
\end{equation}

\begin{figure}[t!] 
	\centering
	\noindent
	\includegraphics[width=14cm]{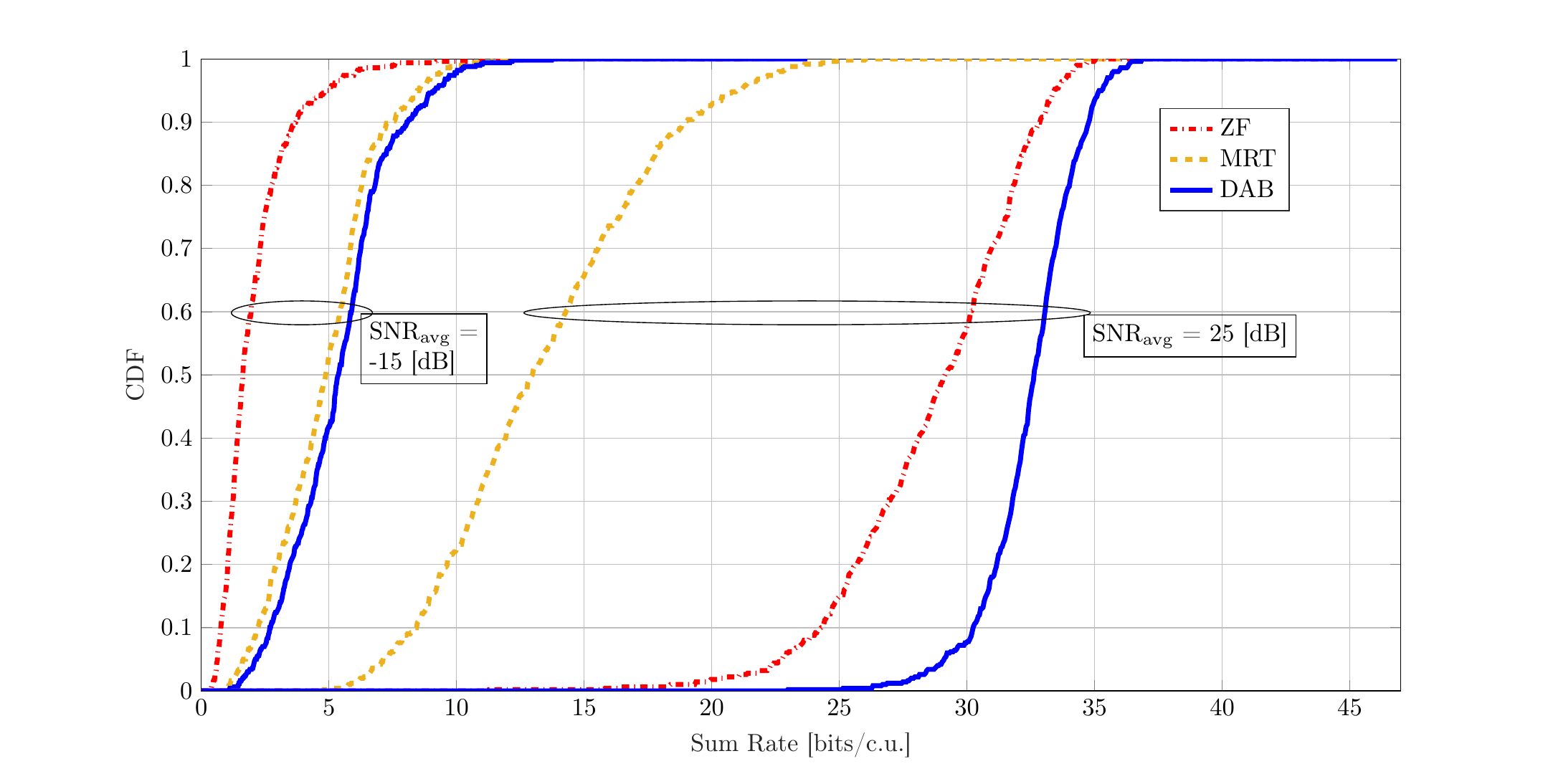}
	\caption{CDF of sum rates with different linear precoders; geometric channel model, $B = 32$ antennas, $U = 4$ users, equal PAs with  $\beta_1^{(b)} = 1$, and $\beta_3^{(b)} = -0.049-0.023j$ for $b = 1, \dots, B$, $\rho_{\text{tot}} = 43$\,dBm, and $\text{SNR}_{\text{avg}} \in \{-15, 25\}$\,dB.}
	\label{CDF_diff_SNR}
\end{figure}

\subsection{Spectral Efficiency: DAB versus Conventional Precoders} \label{subsec:SE}
We compare the spectral efficiency of Algorithm~1, namely DAB, with that of conventional precoding techniques including ZF and MRT.
To this end, we fix the total transmit power to $\rho_{\text{tot}} = 43$\,dBm and characterize the statistics of the achievable sum rate in \eqref{eq:sum_rate} for two different average SNR values, namely $-15$ and $25$\,dB.
In Fig. \ref{CDF_diff_SNR}, the cumulative distribution function (CDF) of the achievable sum rate is plotted for DAB, ZF, and MRT.
In this example, $B = 32$ antennas and $U = 4$ users are considered.
We assume that all the antennas at the transmitter are equipped with equal PAs with $\beta_1^{(b)} = 1$, and $\beta_3^{(b)} = -0.049-0.023j$ for $b = 1, \dots, B$.
We run Algorithm~1 with $20$ initializations and set the number of iterations per initialization to $I = 50$. Moreover, the initial step size is set to $\mu^{(0)} = 0.1$.
From Fig. \ref{CDF_diff_SNR}, it can be seen that the DAB precoding technique outperforms the conventional linear precoders, which do not take into account the distortion introduced by the nonlinear PAs.
The improvement in performance is more significant at high SNR values. This is because, at low SNR (i.e., $\text{SNR}_{\text{avg}} = -15$\,dB), the thermal noise dominates the nonlinear distortion. As a result, dealing with distortion is not necessary and the DAB precoder converges to the MRT precoder (as was also shown in Fig.~\ref{awesome_plot}). 
In the high-SNR regime, however, the performance of both MRT and ZF is limited by the nonlinear distortion. As a result, DAB provides considerable gains over these two conventional precoding techniques.

\begin{figure}[t!] 
	\centering
	\noindent
	\includegraphics[width=14cm]{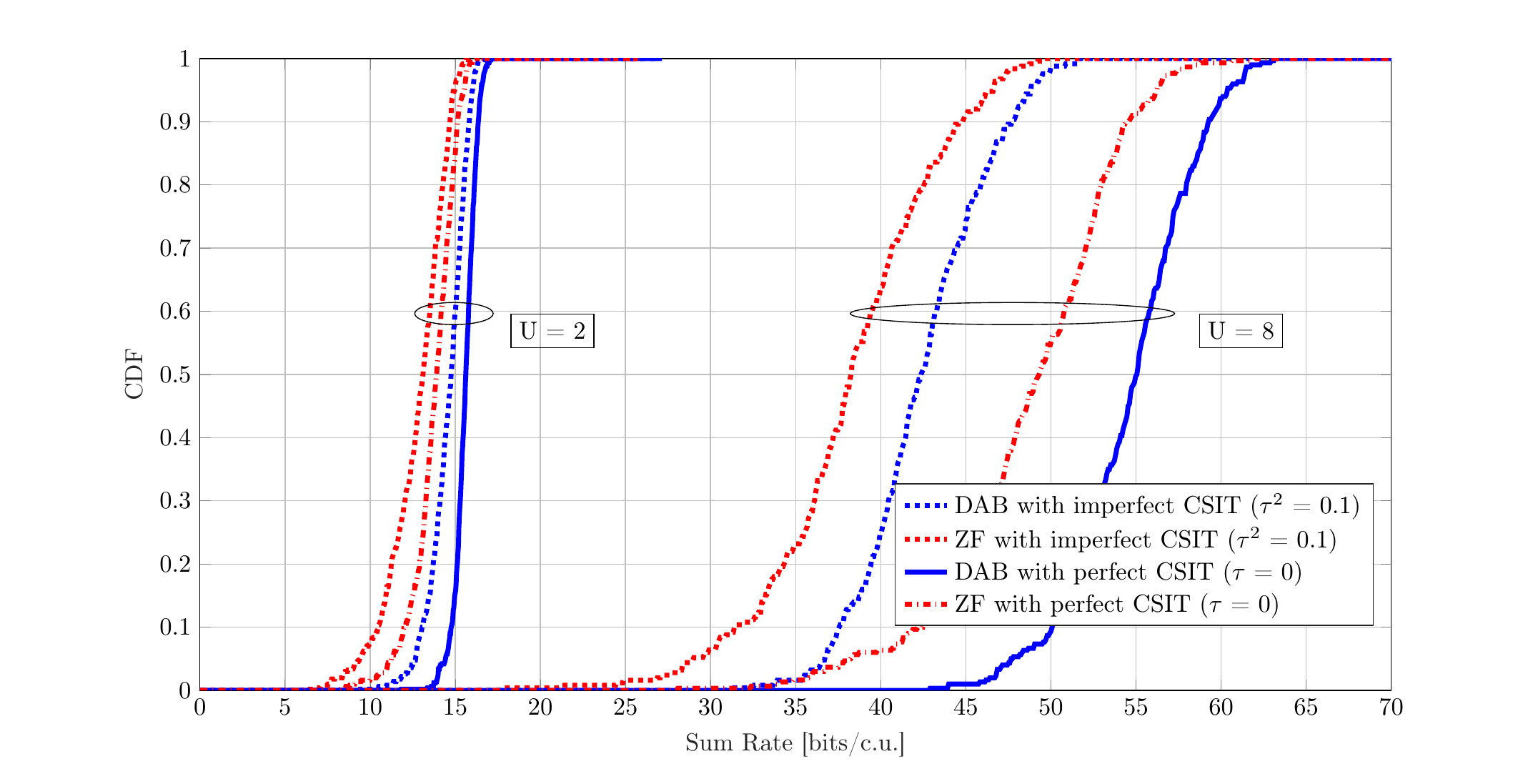}
	\caption{CDF of the sum rate with ZF and DAB precoders with perfect and imperfect CSI at the transmitter; geometric channel model, $B = 32$ antennas, $U \in \{2, 8\}$ users, equal PAs with  $\beta_1^{(b)} = 1$, and $\beta_3^{(b)} = -0.049-0.023j$ for $b = 1, \dots, B$, $\rho_{\text{tot}} = 43$\,dBm, $N_0 = -85$\,dBm (corresponding to $\text{SNR}_{\text{avg}} = 18$\,dB).}
	\label{CDF_with_imp}
\end{figure}

In Fig. \ref{CDF_with_imp}, we consider a setup with $B = 32$ antennas and $U \in \{2, 8\}$ users and illustrate the performance gains achieved by DAB over ZF at $\text{SNR}_{\text{avg}} = 18$\,dB, with both perfect and imperfect CSI at the transmitter (CSIT).
For this purpose, we model the channel estimation error as an additive independent random error term.
Accordingly, the estimated CSIT is modeled by
\begin{equation}
\widehat{\vech}_u = \sqrt{1-\tau^2} \vech_u + \tau \vecv,
\end{equation}
where $\widehat{\vech}_u$ and $\vech_u$ denote the estimated and actual channel corresponding to the $u$th user, respectively, $\tau \in [0, 1]$ is a parameter that reflects the accuracy of channel estimation and the elements of the CSI error $\vecv$ are distributed according to $\mathcal{CN}\left(0, \sigma_{\vech_u}^2\right)$.
Fig. \ref{CDF_with_imp} shows the CDF of the sum rate for ZF and DAB with $\tau = 0$ (perfect CSIT) and $\tau = \sqrt{0.1}$ (imperfect CSIT).
Our results show that once computed with imperfect CSIT, DAB and ZF experience roughly equal performance degradation in terms of achievable sum rate.
For example, at $\text{CDF} = 0.9$, this degradation is about $5\%$ and $20\%$ for both precoding schemes in the scenarios with $U = 2$ and $U = 8$ users, respectively.

\subsection{DAB versus Conventional Precoders in Presence of Power Control} \label{subsec:pwrcntr}
In Figs. \ref{CDF_diff_SNR} and \ref{CDF_with_imp}, we focused on scenarios with fixed total transmit power and demonstrated the advantages of distortion-aware precoding in distortion-limited regimes.
We now consider the case in which the transmitter adopts a power control strategy along with the linear precoding matrix calculation.
In particular, we consider a maximum total transmit power $\rho_{\text{tot},\text{max}}$ and similar to \cite{jee2020precoding}, find precoding matrices that maximize the sum rate under the following power constraint
\begin{equation}\label{eq:max_total_pwr}
\Ex{}{\vecnorm{f(\matP\vecs)}^2} \leq \rho_{\text{tot},\text{max}}.
\end{equation}

\begin{figure}[t!] 
	\centering
	\noindent
	\includegraphics[width=12cm]{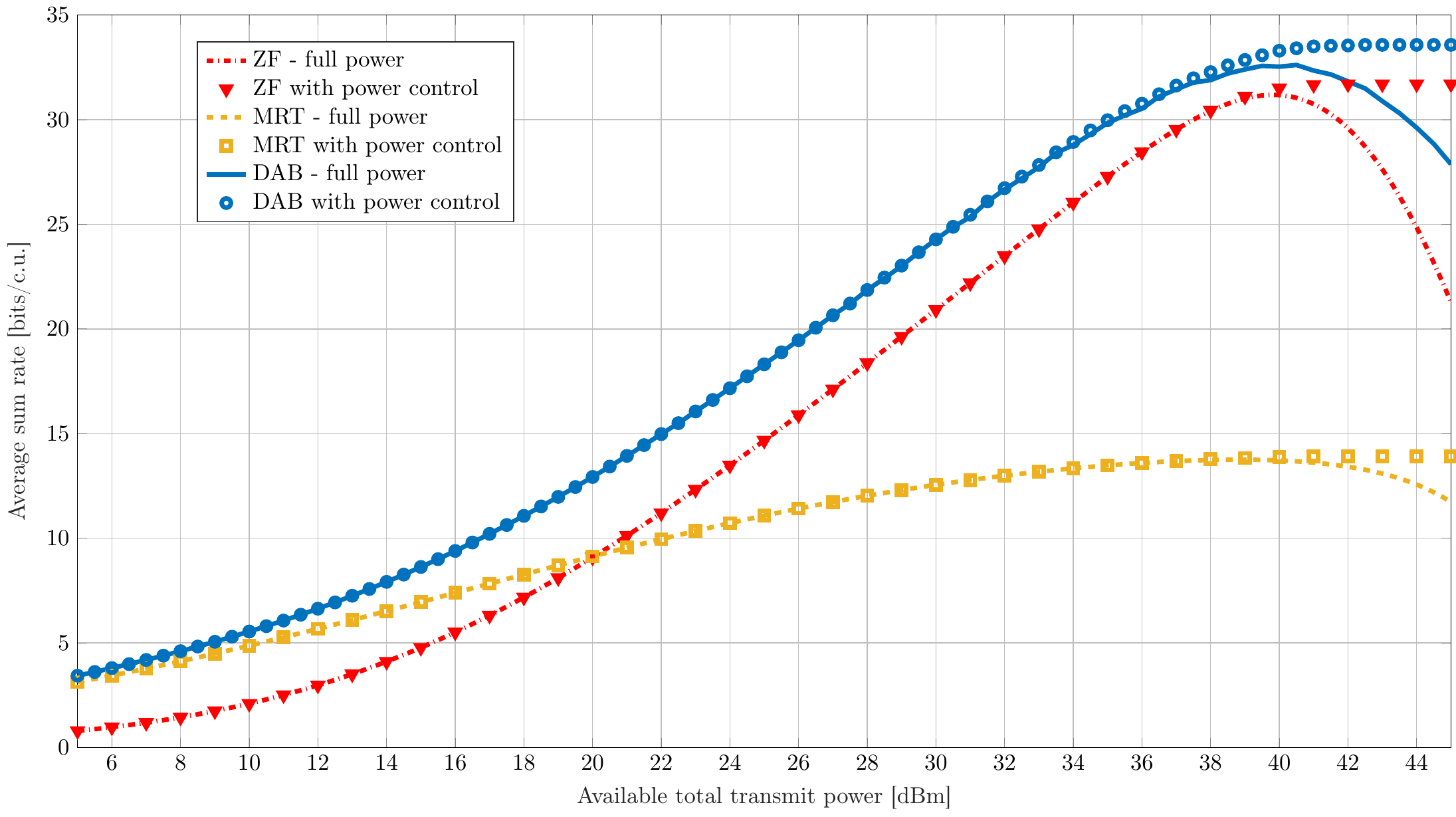}
	\caption{Average sum rate with different linear precoders with and without power control; geometric channel model, $B = 32$ antennas, $U = 4$ users, equal PAs with  $\beta_1^{(b)} = 1$, and $\beta_3^{(b)} = -0.049-0.023j$ for $b = 1, \dots, B$, $N_0 = -85$\,dBm.}
	\label{avsr}
\end{figure}

Fig. \ref{avsr} shows the average sum rate values achieved by MRT, ZF, and DAB precoding techniques under different available total transmit power values for a setup with $B = 32$ antennas and $U = 4$ users.
The channel model and the PA nonlinearity parameters are the same as in Fig.~\ref{CDF_with_imp}.
The full power transmission corresponds to $\rho_{\text{tot}} = \rho_{\text{tot},\text{max}}$ and in the precoding with power control, we calculate the precoding matrices for different values of $\rho_{\text{tot}} \in [0, \rho_{\text{tot},\text{max}}]$ and pick the one that gives the highest sum rate value.
The results in Fig. \ref{avsr} show how DAB performs compared to the conventional linear precoding techniques in different power regimes.
It can be seen that, in the low-power regime, DAB converges to the MRT precoding due to operating in the AWGN-dominated region (similar to the case with $\text{SNR}_{\text{avg}} = -15$\,dB in Fig. \ref{CDF_diff_SNR}).
Once the transmit power is increased, DAB yields improved performance compared to ZF and MRT.
This improved performance follows because the array gain, multiuser interference, and nonlinear distortion are jointly considered in the precoding matrix optimization procedure.
In this example, with full power transmission, MRT, ZF, and DAB reach their maximum average sum rate roughly at $\rho_{\text{tot}} = 38.2$\,dBm, $\rho_{\text{tot}} = 39.7$\,dBm and  $\rho_{\text{tot}} = 40.5$\,dBm, respectively.
With power control, the average sum rate remains constant even when $\rho_{\text{tot},\text{max}}$ is increased further.
Accordingly, DAB provides a constant improvement over ZF and MRT for $\rho_{\text{tot}} \geq 40.5$\,dBm.

\begin{figure}[t!] 
	\centering
	\noindent
	\includegraphics[width=12cm]{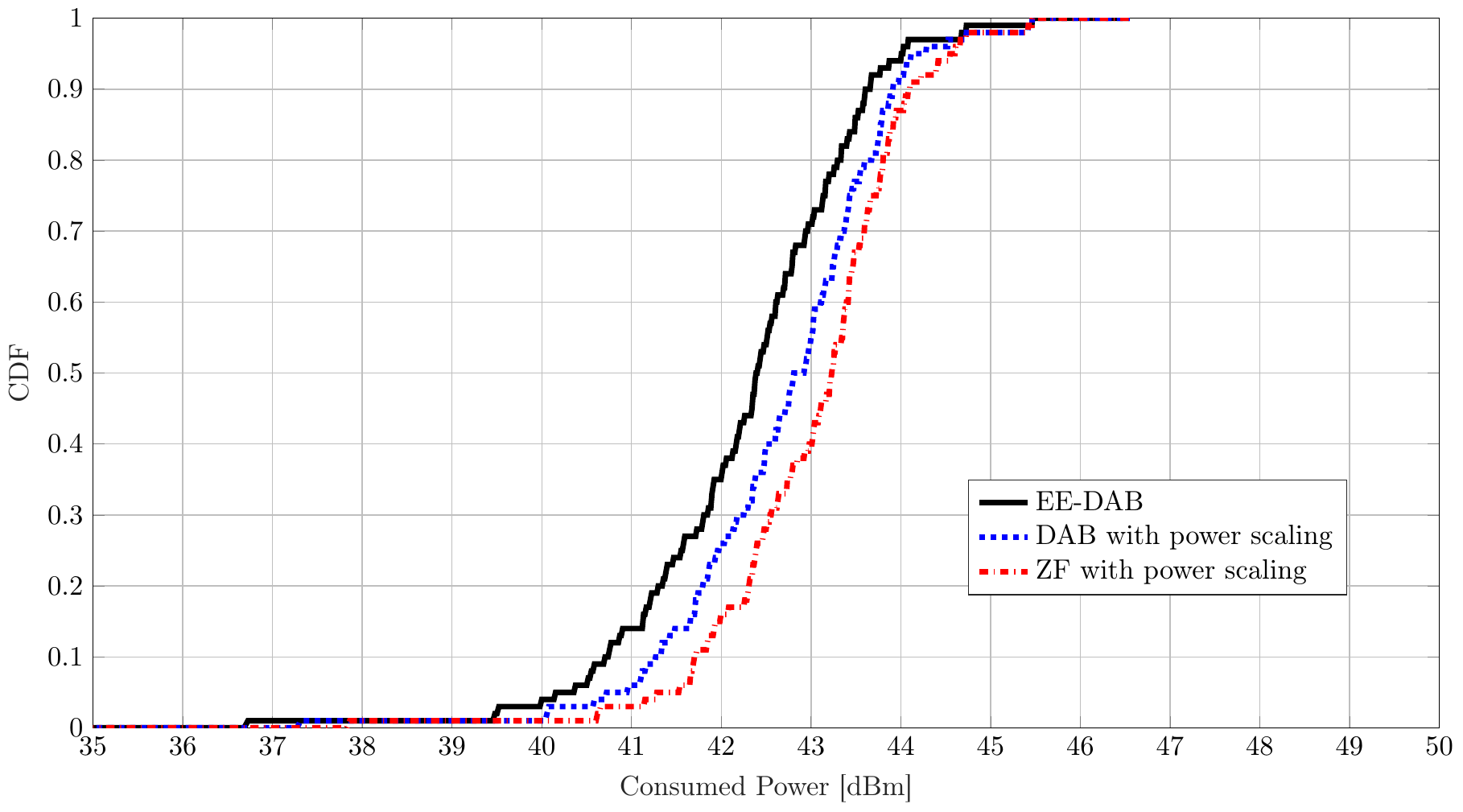}
	\caption{CDF of total base station side consumed power with ZF, DAB and EE-DAB precoders for $R_0 = 35$ (bits/c.u.); geometric channel model, $B = 64$ antennas, $U = 4$ users, equal PAs with  $\beta_1^{(b)} = 1$, and $\beta_3^{(b)} = -0.049-0.023j$ for $b = 1, \dots, B$, $N_0 = -85$\,dBm, power consumption model parameters $\eta_{\text{max,}{b}} = 0.55$ and $\rho_{\text{max,}{b}}= 25$\,dBm.}
	\label{CDF_rho_cons}
\end{figure}

\subsection{Energy Efficiency: EE-DAB versus ZF} \label{subsec:EE}
To evaluate the energy efficiency of EE-DAB precoding, we plot the CDF of the consumed power $\sum_{b = 1}^{B}  \rho_{\text{cons,}{b}}$ for different minimum required sum rate values $R_0$.
We consider a setup with $B = 64$ and $U = 4$ and assume that the channel model and the PA nonlinearity parameters are the same as in Section \ref{subsec:SE}.
The parameters associated with PA power consumption are set to $\eta_{\text{max,}{b}} = 0.55$ and $\rho_{\text{max,}{b}}= 25$\,dBm for all $b = 1, \dots, B$. 
The noise power is $N_0 = -85$\,dBm and the consumed-power values are calculated for the precoding matrices acquired via Algorithm~2 for $R_0 = 35$ bits per channel use (bits/c.u.). 
As benchmarks, we also evaluate the power consumption with the ZF and DAB precoders with a total transmit power normalization such that $R_0$ is achieved.
The resulting CDF is given in Fig. \ref{CDF_rho_cons}.

It can be seen from Fig. \ref{CDF_rho_cons} that, by taking the power consumption into account (in addition to the effect of nonlinear distortion), EE-DAB can achieve a given sum rate with a lower consumed power than ZF and DAB precoding schemes.
This improvement is achieved via optimizing the transmit power in addition to the directivity of the linear signal and distortion using Algorithm~2.

\subsection{Convergence Speed of the Proposed Algorithms}

The improved performance of the DAB and EE-DAB precoders compared to conventional linear precoders is achieved at the cost of increased computational complexity. 
Here we \textit{empirically} study the convergence of Algorithms 1 and 2, which are used to obtain the DAB and EE-DAB precoders, respectively. This provides insight into the number of iterations required to achieve a given performance.

\begin{figure}[t]
	\centering
	\subfloat[DAB, $\rho_\text{tot} = 43$\,dBm.]{\includegraphics[width = .44\textwidth]{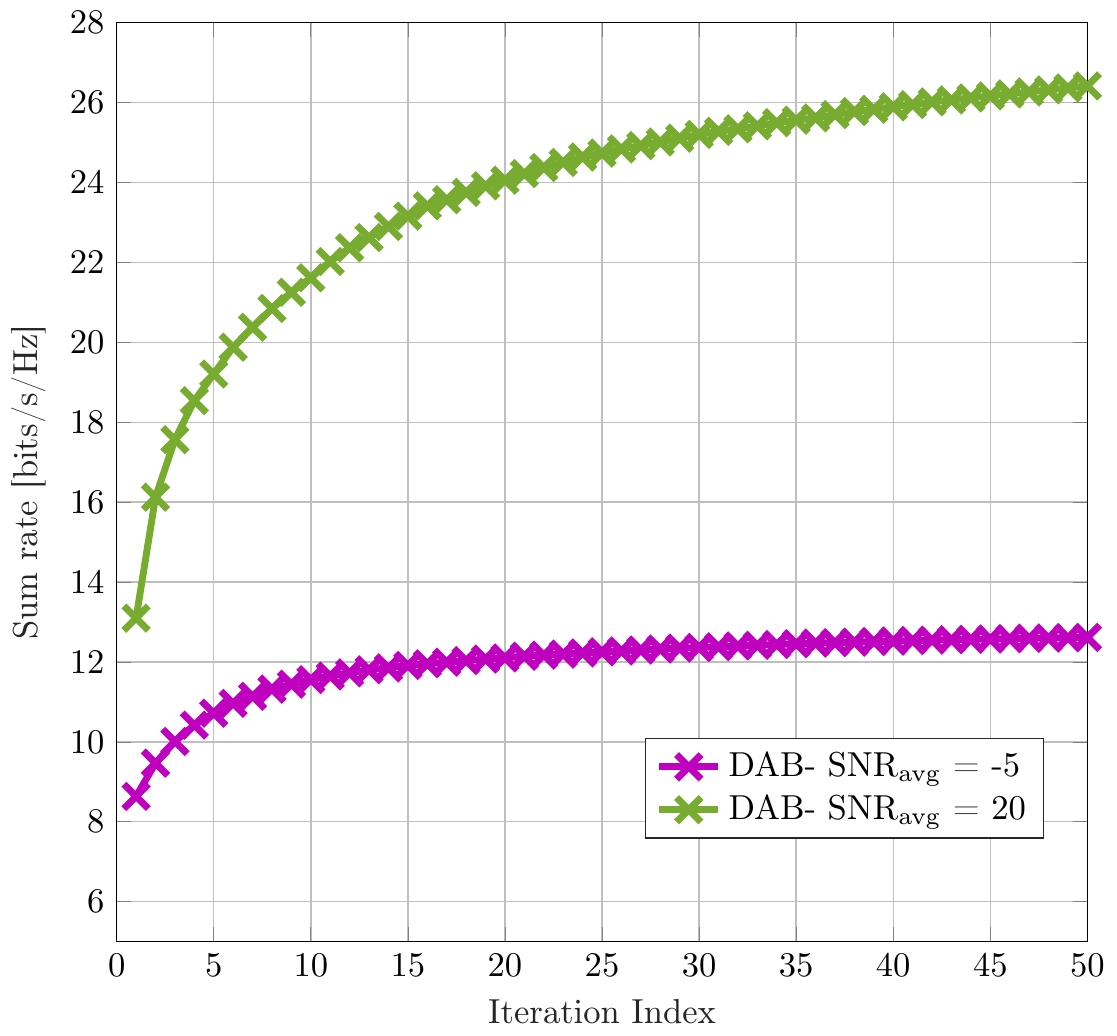}\label{fig:converg_SE}} \quad\quad
	\subfloat[EE-DAB, $\eta_{\text{max,}{b}} = 0.55$ and $\rho_{\text{max,}{b}}= 25$\,dBm for $b = 1, \dots, B$.]{\includegraphics[width = .44\textwidth]{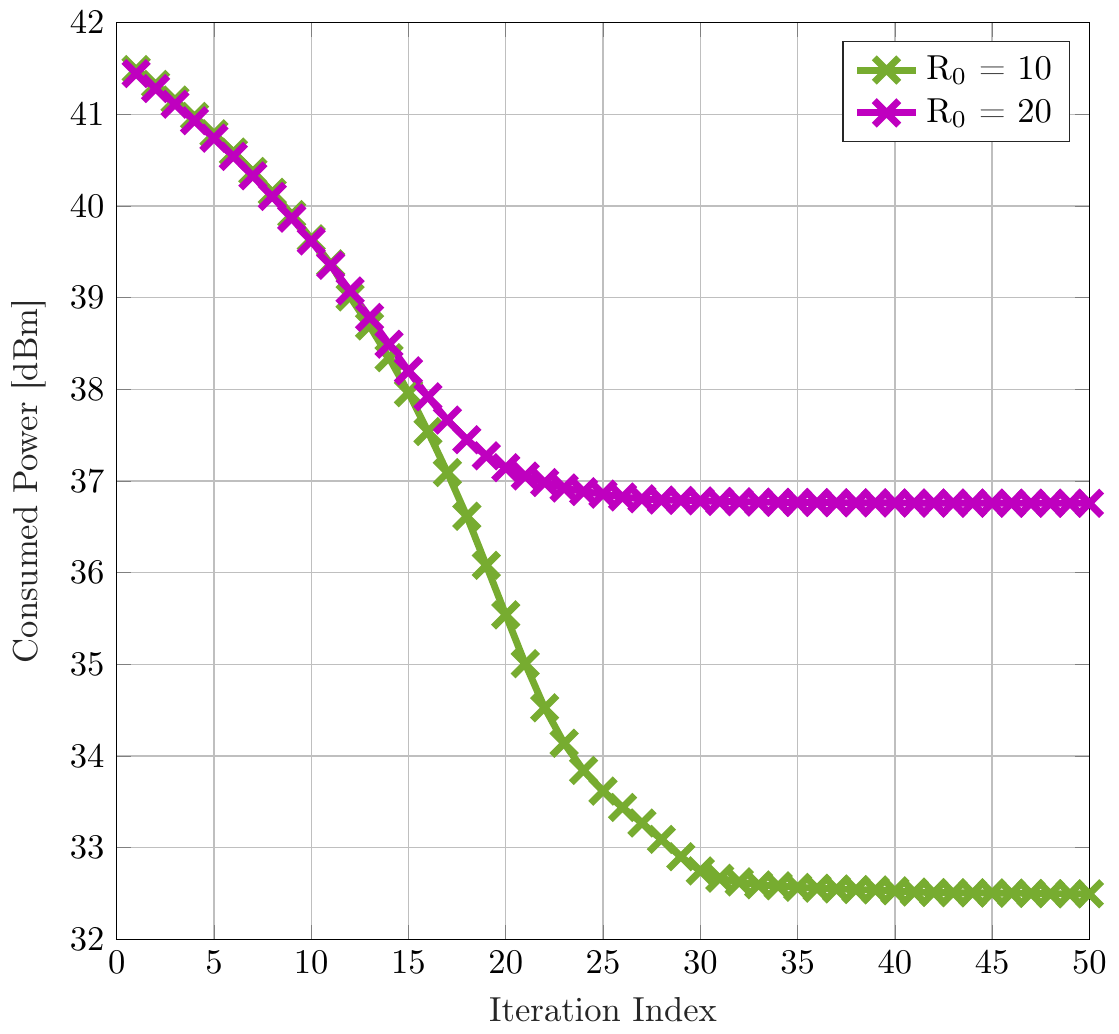}\label{fig:converg_EE}} 
	\caption{Average convergence behavior of Algorithms 1 and 2; geometric channel model, $B=32$ antennas, $U = 4$ users, $\beta_1^{(b)} = 1$, $\beta_3^{(b)} = -0.049-0.023j$ for $b = 1, \dots, B$ and $N_0 = -85$\,dBm. Both algorithms converge in a relatively low number of iterations.}
	\label{fig:convergence}
\end{figure} 

We consider a setup with $B = 32$ and $U = 4$. 
In Fig. \ref{fig:converg_SE}, we show the sum rate achieved by the DAB precoder as a function of the number of iterations. 
It can be seen that the convergence of Algorithm~1 is faster at low SNR compared to high SNR.
Indeed, at low SNR, conventional precoding schemes are as good as DAB precoding (see Fig. \ref{CDF_diff_SNR}).
Therefore, since the MRT and ZF precoding matrices are included in the initialization, fast convergence is expected. 
At high SNR, however, the structure of the optimal precoder is significantly different from that of the MRT and ZF precoders, which results in slower convergence.
Fig. \ref{fig:converg_EE} depicts the convergence behavior of Algorithm~2. 
This demonstrates that with the two proposed algorithms, it is possible to achieve a certain performance in a relatively small number of iterations, keeping the computational complexity within a manageable range.

\subsection{Out-of-Band Emissions}\label{OOB}
In addition to analyses of the in-band distortion caused by the nonlinear PAs, which has been the main focus of this paper, analyses of out-of-band emissions are also of significant practical importance. 
With out-of-band emission, we refer to emissions at frequencies outside the allocated frequency band due to the PA nonlinearities.
These emissions, which are undesirable because they interfere with concurrent transmission in adjacent bands, are strictly regulated in wireless standards.
%

\begin{figure}[!t]
	\centering	
	\includegraphics[width = .45\columnwidth]{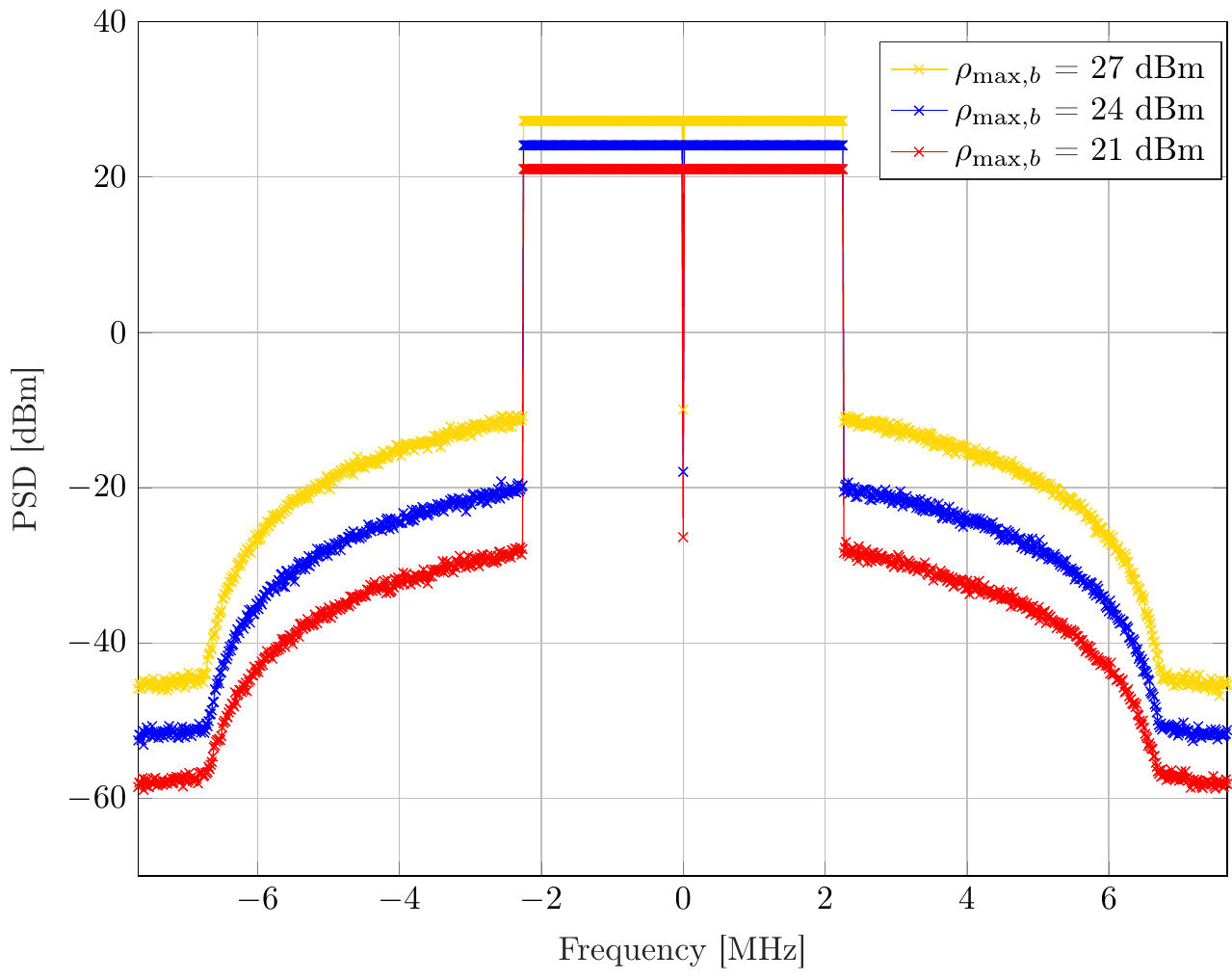}
	\caption{PSD of transmitted signal from the antenna with the highest spectral regrowth with DAB under per-antenna power constraint in \eqref{max_per_antenna}; LoS channel, $B = 16$ antennas, $U = 1$ users, equal PAs as in \eqref{eq:polyn_with_odd}. By controlling $\rho_{\text{max,}{b}}$, we can bound the spectral regrowth within a given mask limit.}
	\label{fig:OOB_emission}
\end{figure}

We evaluate the spectral emissions with DAB precoding under a per-antenna power constraint~\eqref{max_per_antenna} for different $\rho_{\text{max,}{b}}$ values. 
In Fig. \ref{fig:OOB_emission}, we plot the power spectral density (PSD) of the transmitted signal from the antenna with the highest spectral regrowth (worst case) in the array.
We assume a LoS channel with $B = 16$ and $U = 1$ and consider the following nonlinearity model for all PAs in the array
\begin{equation} \label{eq:polyn_with_odd}
f_b(x_b) = \beta_1^{(b)} x_b + \beta_2^{(b)} x_b|x_b| + \beta_3^{(b)} x_b|x_b|^2,
\end{equation}
with $\beta_1^{(b)}= 1$, $\beta_2^{(b)} = -0.02-0.01j$ and $\beta_3^{(b)} = -0.049-0.023j$ for $b =  1, \dots, B$.
The precoding matrix is calculated using Algorithm~1 (while replacing Step 3 with \eqref{eq:update_precoding_matrix_ineq} to satisfy per-antenna transmit power constraints) and applied to an OFDM signal with a one-side bandwidth of $2.25$ MHz.
The total number of subcarriers is $1024$, among which $300$ subcarriers are occupied (the occupied carriers are the first $150$ to the left and to the right of the DC carrier).
The results in Fig.~\ref{fig:OOB_emission} indicate that by controlling $\rho_{\text{max,}{b}}$, one can limit the spectral regrowth outside a given frequency band.
It is indeed possible to generalize the proposed precoder optimization framework to account explicitly for out-of-band emission on top of in-band distortion.
A detailed study of this more general framework is left for future work.

\section{Conclusions} \label{sec:Concl}

We have shown that the choice of precoding matrix has a direct impact on the amount and directivity of distortion emitted by nonlinear power amplifiers in the downlink of massive MIMO systems.
Accordingly, we have introduced precoding optimization algorithms that incorporate knowledge of the nonlinearity characteristics of the PAs as well as their power consumption.
Using numerical examples, we have shown that our proposed solutions can achieve significant improvements in spectral and energy efficiency compared to conventional precoding techniques, i.e., zero-forcing and maximal ratio transmission.
Moreover, by studying out-of-band emission, we have shown that our proposed precoder optimization framework can also be used to limit the spectral regrowth due to nonlinear power amplifiers outside a given frequency band.

\appendices
\section{Closed-Form Expression for $\nabla_\matP R_{\text{sum}}(\matP)$} \label{sec:App_gradient}

We derive a closed-form expression for the gradient $\nabla_\matP R_{\text{sum}}(\matP)$, which facilitates a more accurate evaluation of the update in \eqref{eq:update_precoding_matrix} compared to its approximate numerical evaluation in~\eqref{eq:grd_numerical}.
To this end, let $n_u(\mathbf{P}) = |\mathbf{h}^T_u \mathbf{G}(\mathbf{P}) \mathbf{p}_u|^2$ and $	d_u(\mathbf{P}) = d^\text{\,mui}_u(\matP)  + d^\text{\,dist}_u(\matP)  + N_0$
denote the numerator and denominator of the SINDR in~\eqref{eq:SINR}, respectively, where $d^\text{\,mui}_u(\matP) = \sum_{r \neq u} |\vech^T_u \matG(\matP) \vecp_r|^2$ and $d^\text{\,dist}_u(\matP) = \vech^T_u \matC_{\vece}(\matP) \vech_u^{*}$ is the part of the denominator corresponding to multiuser interference and nonlinear distortion, respectively.
With these definitions, the gradient $\nabla_{\matP} R_\text{sum}(\matP)$ can be written as
\begin{IEEEeqnarray}{rCl}
	\nabla_\matP{R}_\text{sum}(\matP)
	= \sum_{u=1}^U \frac{2 \log_2(e)}{{d_u^2(\matP)(1 + n_u(\matP)/d_u(\matP))}} 
	 \left( d_u(\matP)\frac{\partial n_u(\matP)}{\partial\matP^*}- n_u(\matP) \frac{\partial d_u(\matP)}{\partial\matP^*}\right), \label{eq:gradient_of_sum_rate} \IEEEeqnarraynumspace
\end{IEEEeqnarray}
where $\partial n_u(\matP) / \partial\matP^* = \splitatcommas{[\partial{n_u}(\matP)/\partial\vecp_1^*, \dots, \partial{n_u}(\matP)/\partial\vecp_u^*]}$ and $\partial d_u(\matP) / \partial\matP^* = \splitatcommas{[\partial{d_u}(\matP)/\partial\vecp_1^*, \dots, \partial{d_u}(\matP)/\partial\vecp_U^*]}$. 
Hence, to compute the gradient $\nabla_{\mathbf{P}} R_\text{sum}(\mathbf{P})$, we need to compute the derivatives of $n_u(\matP)$ and $d_u(\matP)$ for $u = 1,\dots,U$.
We consider the third-order nonlinearity, i.e., $K = 1$ in \eqref{eq:polyn} and accordingly $\matG(\matP)$ and $\matC_{\vece}(\matP)$ are as in \eqref{eq:BG_3rd_order} and \eqref{eq:Ce_3rd_order}, respectively.
Starting with the numerator, it can be shown~that the derivative with respect to $\vecp_{u'}^*$ can be written~as
\begin{IEEEeqnarray}{rCl} \label{eq:deriv_numerator}
	\frac{\partial{n_u}(\matP)}{\partial\vecp_{u'}^*} 
	&=& \left(\mathbf{\Gamma}_{u}(\matP) \mathds{1}_{\{{u'} = u\}}(u') + \mathbf{\Upsilon}_{u,u'}(\matP)\right)\vecp_{u'}, 
\end{IEEEeqnarray}
for $u' = 1,\dots,U$, where $\mathbf{\Gamma}_{u}(\matP) \in \opC^{B \times B}$ is given by
\begin{align}\label{num_deriv1}
	\mathbf{\Gamma}_{u}(\matP) 
	=&  \matA_1^*  \vech_u^*\vech_u^T \matA_1
	 + 2 \left( \matA_1^*  \vech_u^*\vech_u^T  \matA_3 \text{diag}(\matP\matP^H) \right.
	 + \left.  \text{diag}(\vecp_{u}\vecp_{u}^H \matA_1^* \vech_u^*\vech_u^T \matA_3 )  \right) \nonumber\\
	 &+ 2 \left( \text{diag}(\matP\matP^H) \matA_3^*  \vech_u^*\vech_u^T  \matA_1  \right.
	 + \left.  \text{diag}(\matA_3^* \vech_u^*\vech_u^T \matA_1 \vecp_{u}\vecp_{u}^H) \right)  \nonumber\\
	 &+ 4 \big( \text{diag}(\matP\matP^H) \matA_3^* \vech_u^*\vech_u^T \matA_3^* \text{diag}(\matP\matP^H) + \text{diag}(\matA_3^*  \vech_u^*\vech_u^T \matA_3 \text{diag}(\matP\matP^H) \vecp_{u}\vecp_{u}^H \nonumber \\&+ \vecp_{u}\vecp_{u}^H \text{diag}(\matP\matP^H) \matA_3^*  \vech_u^*\vech_u^T \matA_3 ) \big).
\end{align}
Furthermore, $\mathbf{\Upsilon}_{u}(\matP) \in \opC^{B \times B}$ can be expressed as
\begin{align}\label{num_deriv2}
	\mathbf{\Upsilon}_{u,u'}(\matP)
	=& 2 (\text{diag}( \vecp_{u'}\vecp_{u'}^H \matA_1^* \vech_u^*\vech_u^T \matA_3 ) 
	 +  \text{diag}(\matA_1^* \vech_u^*\vech_u^T  \matA_3 \vecp_{u'}\vecp_{u'}^H)) \nonumber \\
    +&  4 ( \text{diag}(\matA_3^* \vech_u^*\vech_u^T \matA_3 \text{diag}(\matP\matP^H) \vecp_{u'}\vecp_{u'}^H )   +  \text{diag}( \vecp_{u'}\vecp_{u'}^H \text{diag}(\matP\matP^H) \matA_3^* \vech_u^*\vech_u^T \matA_3) ).
\end{align}
The derivative with respect to $\vecp_{u'}^*$ of the denominator can be written as
\begin{IEEEeqnarray}{rCl} \label{eq:deriv_denom}
	\frac{\partial{d_u}(\matP)}{\partial\vecp_{u'}^*} &=& 	\frac{\partial{d^\text{\,mui}_u}(\matP)}{\partial\vecp_{u'}^*} + \frac{\partial{d^\text{\,dist}_u}(\matP)}{\partial\vecp_{u'}^*},
\end{IEEEeqnarray}
where the derivative of the multiuser interference term in the denominator is given~by
\begin{IEEEeqnarray}{rCl} \label{eq:deriv_denom_mui}
	\frac{\partial{d^\text{\,mui}_u}(\matP)}{\partial\vecp_{u'}^*}
	&=& \left(\mathbf{\Gamma}_{u}(\matP) \mathds{1}_{\{{u'} \neq u\}}(u') + \sum_{r \neq u} \mathbf{\Upsilon}_{u,r}(\matP)\right)\!\vecp_{u'} \IEEEeqnarraynumspace
\end{IEEEeqnarray}
for $u' = 1, \dots, U$. Furthermore, the $b$th entry of the derivative of the nonlinear distortion term in the denominator is given by
\begin{IEEEeqnarray}{rCl} \label{eq:deriv_denom_dist}
	\frac{\partial{d^\text{\,dist}_u}(\matP)}{\partial{p^*_{u',b}}}
	= 2 \abs{\beta_3^{(b)}}^2 \! \left( 2 h^*_{u,b} \!\! \sum_{b'=1}^B \! h_{u,b'} p_{u',b'} \left[ \abs{\matP\matP^H}^2\right]_{b',b} \right. \quad  + \left. h_{u,b} \!\! \sum_{b'=1}^B \! h_{u,b'}^* p_{u',b'}\left[ \left(\matP\matP^H\right)^2\right]_{b,b'} \right) \IEEEeqnarraynumspace 
\end{IEEEeqnarray}
for $u' = 1, \dots, U$ and $b = 1, \dots, B$, where $h_{b,u} = [\vech_u]_b$, $p_{b,u} = [\vecp_u]_b$ and $\beta_3^{(b)}$ is the coefficient corresponding to the $b$th PA in the array.
Finally, by inserting \eqref{eq:deriv_numerator}, \eqref{eq:deriv_denom}, \eqref{eq:deriv_denom_mui}, and \eqref{eq:deriv_denom_dist} into \eqref{eq:gradient_of_sum_rate}, we obtain a closed-form expression for the gradient $\nabla_{\mathbf{P}} 	R_\text{sum}(\mathbf{P})$.

\bibliographystyle{IEEEtran}
\bibliography{IEEEabrv,confs-jrnls,publishers,svenbib}
\balance
\end{document}

%% file: vmr-symbols-vecbold.tex
\usepackage{amssymb}
\usepackage{amsfonts}
\usepackage{mathrsfs}
\usepackage{xspace}
\usepackage{bm}
\usepackage{upgreek}

\newcommand{\safemath}[2]{\newcommand{#1}{\ensuremath{#2}\xspace}}

\safemath{\bma}{\mathbf{a}}
\safemath{\bmb}{\mathbf{b}}
\safemath{\bmc}{\mathbf{c}}
\safemath{\bmd}{\mathbf{d}}
\safemath{\bme}{\mathbf{e}}
\safemath{\bmf}{\mathbf{f}}
\safemath{\bmg}{\mathbf{g}}
\safemath{\bmh}{\mathbf{h}}
\safemath{\bmi}{\mathbf{i}}
\safemath{\bmj}{\mathbf{j}}
\safemath{\bmk}{\mathbf{k}}
\safemath{\bml}{\mathbf{l}}
\safemath{\bmm}{\mathbf{m}}
\safemath{\bmn}{\mathbf{n}}
\safemath{\bmo}{\mathbf{o}}
\safemath{\bmp}{\mathbf{p}}
\safemath{\bmq}{\mathbf{q}}
\safemath{\bmr}{\mathbf{r}}
\safemath{\bms}{\mathbf{s}}
\safemath{\bmt}{\mathbf{t}}
\safemath{\bmu}{\mathbf{u}}
\safemath{\bmv}{\mathbf{v}}
\safemath{\bmw}{\mathbf{w}}
\safemath{\bmx}{\mathbf{x}}
\safemath{\bmy}{\mathbf{y}}
\safemath{\bmz}{\mathbf{z}}
\safemath{\bmzero}{\mathbf{0}}
\safemath{\bmone}{\mathbf{1}}

\bmdefine{\biad}{a}
\bmdefine{\bibd}{b}
\bmdefine{\bicd}{c}
\bmdefine{\bidd}{d}
\bmdefine{\bied}{e}
\bmdefine{\bifd}{f}
\bmdefine{\bigd}{g}
\bmdefine{\bihd}{h}
\bmdefine{\biid}{i}
\bmdefine{\bijd}{j}
\bmdefine{\bikd}{k}
\bmdefine{\bild}{l}
\bmdefine{\bimd}{m}
\bmdefine{\bind}{n}
\bmdefine{\biod}{o}
\bmdefine{\bipd}{p}
\bmdefine{\biqd}{q}
\bmdefine{\bird}{r}
\bmdefine{\bisd}{s}
\bmdefine{\bitd}{t}
\bmdefine{\biud}{u}
\bmdefine{\bivd}{v}
\bmdefine{\biwd}{w}
\bmdefine{\bixd}{x}
\bmdefine{\biyd}{y}
\bmdefine{\bizd}{z}

\bmdefine{\bixid}{\xi}
\bmdefine{\bilambdad}{\lambda}
\bmdefine{\bimud}{\mu}
\bmdefine{\bithetad}{\theta}
\bmdefine{\biphid}{\phi}
\bmdefine{\bideltad}{\delta}

\safemath{\bmia}{\biad}
\safemath{\bmib}{\bibd}
\safemath{\bmic}{\bicd}
\safemath{\bmid}{\bidd}
\safemath{\bmie}{\bied}
\safemath{\bmif}{\bifd}
\safemath{\bmig}{\bigd}
\safemath{\bmih}{\bihd}
\safemath{\bmii}{\biid}
\safemath{\bmij}{\bijd}
\safemath{\bmik}{\bikd}
\safemath{\bmil}{\bild}
\safemath{\bmim}{\bimd}
\safemath{\bmin}{\bind}
\safemath{\bmio}{\biod}
\safemath{\bmip}{\bipd}
\safemath{\bmiq}{\biqd}
\safemath{\bmir}{\bird}
\safemath{\bmis}{\bisd}
\safemath{\bmit}{\bitd}
\safemath{\bmiu}{\biud}
\safemath{\bmiv}{\bivd}
\safemath{\bmiw}{\biwd}
\safemath{\bmix}{\bixd}
\safemath{\bmiy}{\biyd}
\safemath{\bmiz}{\bizd}

\safemath{\bmxi}{\bixid}
\safemath{\bmlambda}{\bilambdad}
\safemath{\bmmu}{\bimud}
\safemath{\bmtheta}{\bithetad}
\safemath{\bmphi}{\biphid}
\safemath{\bmdelta}{\bideltad}

\safemath{\bA}{\mathbf{A}}
\safemath{\bB}{\mathbf{B}}
\safemath{\bC}{\mathbf{C}}
\safemath{\bD}{\mathbf{D}}
\safemath{\bE}{\mathbf{E}}
\safemath{\bF}{\mathbf{F}}
\safemath{\bG}{\mathbf{G}}
\safemath{\bH}{\mathbf{H}}
\safemath{\bI}{\mathbf{I}}
\safemath{\bJ}{\mathbf{J}}
\safemath{\bK}{\mathbf{K}}
\safemath{\bL}{\mathbf{L}}
\safemath{\bM}{\mathbf{M}}
\safemath{\bN}{\mathbf{N}}
\safemath{\bO}{\mathbf{O}}
\safemath{\bP}{\mathbf{P}}
\safemath{\bQ}{\mathbf{Q}}
\safemath{\bR}{\mathbf{R}}
\safemath{\bS}{\mathbf{S}}
\safemath{\bT}{\mathbf{T}}
\safemath{\bU}{\mathbf{U}}
\safemath{\bV}{\mathbf{V}}
\safemath{\bW}{\mathbf{W}}
\safemath{\bX}{\mathbf{X}}
\safemath{\bY}{\mathbf{Y}}
\safemath{\bZ}{\mathbf{Z}}

\safemath{\bZero}{\mathbf{0}}
\safemath{\bOne}{\mathbf{1}}
\safemath{\bDelta}{\mathbf{\Delta}}
\safemath{\bLambda}{\mathbf{\UpLambda}}
\safemath{\bPhi}{\mathbf{\Upphi}}
\safemath{\bSigma}{\mathbf{\Upsigma}}
\safemath{\bOmega}{\mathbf{\Upomega}}
\safemath{\bTheta}{\mathbf{\Uptheta}}

\bmdefine{\biAd}{A}
\bmdefine{\biBd}{B}
\bmdefine{\biCd}{C}
\bmdefine{\biDd}{D}
\bmdefine{\biEd}{E}
\bmdefine{\biFd}{F}
\bmdefine{\biGd}{G}
\bmdefine{\biHd}{H}
\bmdefine{\biId}{I}
\bmdefine{\biJd}{J}
\bmdefine{\biKd}{K}
\bmdefine{\biLd}{L}
\bmdefine{\biMd}{M}
\bmdefine{\biOd}{N}
\bmdefine{\biPd}{O}
\bmdefine{\biQd}{P}
\bmdefine{\biRd}{R}
\bmdefine{\biSd}{S}
\bmdefine{\biTd}{T}
\bmdefine{\biUd}{U}
\bmdefine{\biVd}{V}
\bmdefine{\biWd}{W}
\bmdefine{\biXd}{X}
\bmdefine{\biYd}{Y}
\bmdefine{\biZd}{Z}

\bmdefine{\biDelta}{\Delta}
\bmdefine{\biLambda}{\Lambda}
\bmdefine{\biPhi}{\Phi}
\bmdefine{\biSigma}{\Sigma}
\bmdefine{\biOmega}{\Omega}
\bmdefine{\biTheta}{\Theta}

\safemath{\bimA}{\biAd}
\safemath{\bimB}{\biBd}
\safemath{\bimC}{\biCd}
\safemath{\bimD}{\biDd}
\safemath{\bimE}{\biEd}
\safemath{\bimF}{\biFd}
\safemath{\bimG}{\biGd}
\safemath{\bimH}{\biHd}
\safemath{\bimI}{\biId}
\safemath{\bimJ}{\biJd}
\safemath{\bimK}{\biKd}
\safemath{\bimL}{\biLd}
\safemath{\bimM}{\biMd}
\safemath{\bimN}{\biNd}
\safemath{\bimO}{\biOd}
\safemath{\bimP}{\biPd}
\safemath{\bimQ}{\biQd}
\safemath{\bimR}{\biRd}
\safemath{\bimS}{\biSd}
\safemath{\bimT}{\biTd}
\safemath{\bimU}{\biUd}
\safemath{\bimV}{\biVd}
\safemath{\bimW}{\biWd}
\safemath{\bimX}{\biXd}
\safemath{\bimY}{\biYd}
\safemath{\bimZ}{\biZd}

\safemath{\bimDelta}{\biDelta}
\safemath{\bimLambda}{\biLambda}
\safemath{\bimPhi}{\biPhi}
\safemath{\bimSigma}{\biSigma}
\safemath{\bimOmega}{\biOmega}
\safemath{\bimTheta}{\biTheta}

\safemath{\setA}{\mathcal{A}}
\safemath{\setB}{\mathcal{B}}
\safemath{\setC}{\mathcal{C}}
\safemath{\setD}{\mathcal{D}}
\safemath{\setE}{\mathcal{E}}
\safemath{\setF}{\mathcal{F}}
\safemath{\setG}{\mathcal{G}}
\safemath{\setH}{\mathcal{H}}
\safemath{\setI}{\mathcal{I}}
\safemath{\setJ}{\mathcal{J}}
\safemath{\setK}{\mathcal{K}}
\safemath{\setL}{\mathcal{L}}
\safemath{\setM}{\mathcal{M}}
\safemath{\setN}{\mathcal{N}}
\safemath{\setO}{\mathcal{O}}
\safemath{\setP}{\mathcal{P}}
\safemath{\setQ}{\mathcal{Q}}
\safemath{\setR}{\mathcal{R}}
\safemath{\setS}{\mathcal{S}}
\safemath{\setT}{\mathcal{T}}
\safemath{\setU}{\mathcal{U}}
\safemath{\setV}{\mathcal{V}}
\safemath{\setW}{\mathcal{W}}
\safemath{\setX}{\mathcal{X}}
\safemath{\setY}{\mathcal{Y}}
\safemath{\setZ}{\mathcal{Z}}
\safemath{\emptySet}{\varnothing}

\safemath{\colA}{\mathscr{A}}
\safemath{\colB}{\mathscr{B}}
\safemath{\colC}{\mathscr{C}}
\safemath{\colD}{\mathscr{D}}
\safemath{\colE}{\mathscr{E}}
\safemath{\colF}{\mathscr{F}}
\safemath{\colG}{\mathscr{G}}
\safemath{\colH}{\mathscr{H}}
\safemath{\colI}{\mathscr{I}}
\safemath{\colJ}{\mathscr{J}}
\safemath{\colK}{\mathscr{K}}
\safemath{\colL}{\mathscr{L}}
\safemath{\colM}{\mathscr{M}}
\safemath{\colN}{\mathscr{N}}
\safemath{\colO}{\mathscr{O}}
\safemath{\colP}{\mathscr{P}}
\safemath{\colQ}{\mathscr{Q}}
\safemath{\colR}{\mathscr{R}}
\safemath{\colS}{\mathscr{S}}
\safemath{\colT}{\mathscr{T}}
\safemath{\colU}{\mathscr{U}}
\safemath{\colV}{\mathscr{V}}
\safemath{\colW}{\mathscr{W}}
\safemath{\colX}{\mathscr{X}}
\safemath{\colY}{\mathscr{Y}}
\safemath{\colZ}{\mathscr{Z}}

\safemath{\opA}{\mathbb{A}}
\safemath{\opB}{\mathbb{B}}
\safemath{\opC}{\mathbb{C}}
\safemath{\opD}{\mathbb{D}}
\safemath{\opE}{\mathbb{E}}
\safemath{\opF}{\mathbb{F}}
\safemath{\opG}{\mathbb{G}}
\safemath{\opH}{\mathbb{H}}
\safemath{\opI}{\mathbb{I}}
\safemath{\opJ}{\mathbb{J}}
\safemath{\opK}{\mathbb{K}}
\safemath{\opL}{\mathbb{L}}
\safemath{\opM}{\mathbb{M}}
\safemath{\opN}{\mathbb{N}}
\safemath{\opO}{\mathbb{O}}
\safemath{\opP}{\mathbb{P}}
\safemath{\opQ}{\mathbb{Q}}
\safemath{\opR}{\mathbb{R}}
\safemath{\opS}{\mathbb{S}}
\safemath{\opT}{\mathbb{T}}
\safemath{\opU}{\mathbb{U}}
\safemath{\opV}{\mathbb{V}}
\safemath{\opW}{\mathbb{W}}
\safemath{\opX}{\mathbb{X}}
\safemath{\opY}{\mathbb{Y}}
\safemath{\opZ}{\mathbb{Z}}
\safemath{\opZero}{\mathbb{O}}
\safemath{\identityop}{\opI}

\safemath{\veca}{\bma}
\safemath{\vecb}{\bmb}
\safemath{\vecc}{\bmc}
\safemath{\vecd}{\bmd}
\safemath{\vece}{\bme}
\safemath{\vecf}{\bmf}
\safemath{\vecg}{\bmg}
\safemath{\vech}{\bmh}
\safemath{\veci}{\bmi}
\safemath{\vecj}{\bmj}
\safemath{\veck}{\bmk}
\safemath{\vecl}{\bml}
\safemath{\vecm}{\bmm}
\safemath{\vecn}{\bmn}
\safemath{\veco}{\bmo}
\safemath{\vecp}{\bmp}
\safemath{\vecq}{\bmq}
\safemath{\vecr}{\bmr}
\safemath{\vecs}{\bms}
\safemath{\vect}{\bmt}
\safemath{\vecu}{\bmu}
\safemath{\vecv}{\bmv}
\safemath{\vecw}{\bmw}
\safemath{\vecx}{\bmx}
\safemath{\vecy}{\bmy}
\safemath{\vecz}{\bmz}

\safemath{\veczero}{\bmzero}
\safemath{\vecone}{\bmone}
\safemath{\vecxi}{\bmxi}
\safemath{\veclambda}{\bmlambda}
\safemath{\vecmu}{\bmmu}
\safemath{\vectheta}{\bmtheta}
\safemath{\vecphi}{\bmphi}
\safemath{\vecdelta}{\bmdelta}

\safemath{\matA}{\bA}
\safemath{\matB}{\bB}
\safemath{\matC}{\bC}
\safemath{\matD}{\bD}
\safemath{\matE}{\bE}
\safemath{\matF}{\bF}
\safemath{\matG}{\bG}
\safemath{\matH}{\bH}
\safemath{\matI}{\bI}
\safemath{\matJ}{\bJ}
\safemath{\matK}{\bK}
\safemath{\matL}{\bL}
\safemath{\matM}{\bM}
\safemath{\matN}{\bN}
\safemath{\matO}{\bO}
\safemath{\matP}{\bP}
\safemath{\matQ}{\bQ}
\safemath{\matR}{\bR}
\safemath{\matS}{\bS}
\safemath{\matT}{\bT}
\safemath{\matU}{\bU}
\safemath{\matV}{\bV}
\safemath{\matW}{\bW}
\safemath{\matX}{\bX}
\safemath{\matY}{\bY}
\safemath{\matZ}{\bZ}
\safemath{\matzero}{\bmzero}

\safemath{\matDelta}{\bDelta}
\safemath{\matLambda}{\bLambda}
\safemath{\matPhi}{\bPhi}
\safemath{\matSigma}{\bSigma}
\safemath{\matOmega}{\bOmega}
\safemath{\matTheta}{\bTheta}

\safemath{\matidentity}{\matI}
\safemath{\matone}{\matO}

\safemath{\rnda}{A}
\safemath{\rndb}{B}
\safemath{\rndc}{C}
\safemath{\rndd}{D}
\safemath{\rnde}{E}
\safemath{\rndf}{F}
\safemath{\rndg}{G}
\safemath{\rndh}{H}
\safemath{\rndi}{I}
\safemath{\rndj}{J}
\safemath{\rndk}{K}
\safemath{\rndl}{L}
\safemath{\rndm}{M}
\safemath{\rndn}{N}
\safemath{\rndo}{O}
\safemath{\rndp}{P}
\safemath{\rndq}{Q}
\safemath{\rndr}{R}
\safemath{\rnds}{S}
\safemath{\rndt}{T}
\safemath{\rndu}{U}
\safemath{\rndv}{V}
\safemath{\rndw}{W}
\safemath{\rndx}{X}
\safemath{\rndy}{Y}
\safemath{\rndz}{Z}

\safemath{\rveca}{\bimA}
\safemath{\rvecb}{\bimB}
\safemath{\rvecc}{\bimC}
\safemath{\rvecd}{\bimD}
\safemath{\rvece}{\bimE}
\safemath{\rvecf}{\bimF}
\safemath{\rvecg}{\bimG}
\safemath{\rvech}{\bimH}
\safemath{\rveci}{\bimI}
\safemath{\rvecj}{\bimJ}
\safemath{\rveck}{\bimK}
\safemath{\rvecl}{\bimL}
\safemath{\rvecm}{\bimM}
\safemath{\rvecn}{\bimN}
\safemath{\rveco}{\bomO}
\safemath{\rvecp}{\bimP}
\safemath{\rvecq}{\bimQ}
\safemath{\rvecr}{\bimR}
\safemath{\rvecs}{\bimS}
\safemath{\rvect}{\bimT}
\safemath{\rvecu}{\bimU}
\safemath{\rvecv}{\bimV}
\safemath{\rvecw}{\bimW}
\safemath{\rvecx}{\bimX}
\safemath{\rvecy}{\bimY}
\safemath{\rvecz}{\bimZ}

\safemath{\rvecxi}{\bmxi}
\safemath{\rveclambda}{\bmlambda}
\safemath{\rvecmu}{\bmmu}
\safemath{\rvectheta}{\bmtheta}
\safemath{\rvecphi}{\bmphi}

\safemath{\rmatA}{\bimA}
\safemath{\rmatB}{\bimB}
\safemath{\rmatC}{\bimC}
\safemath{\rmatD}{\bimD}
\safemath{\rmatE}{\bimE}
\safemath{\rmatF}{\bimF}
\safemath{\rmatG}{\bimG}
\safemath{\rmatH}{\bimH}
\safemath{\rmatI}{\bimI}
\safemath{\rmatJ}{\bimJ}
\safemath{\rmatK}{\bimK}
\safemath{\rmatL}{\bimL}
\safemath{\rmatM}{\bimM}
\safemath{\rmatN}{\bimN}
\safemath{\rmatO}{\bimO}
\safemath{\rmatP}{\bimP}
\safemath{\rmatQ}{\bimQ}
\safemath{\rmatR}{\bimR}
\safemath{\rmatS}{\bimS}
\safemath{\rmatT}{\bimT}
\safemath{\rmatU}{\bimU}
\safemath{\rmatV}{\bimV}
\safemath{\rmatW}{\bimW}
\safemath{\rmatX}{\bimX}
\safemath{\rmatY}{\bimY}
\safemath{\rmatZ}{\bimZ}

\safemath{\rmatDelta}{\bimDelta}
\safemath{\rmatLambda}{\bimLambda}
\safemath{\rmatPhi}{\bimPhi}
\safemath{\rmatSigma}{\bimSigma}
\safemath{\rmatOmega}{\bimOmega}
\safemath{\rmatTheta}{\bimTheta}

%% file: standard-macros.tex
\usepackage{amssymb}
\usepackage{amsfonts}
\usepackage{mathrsfs}
\usepackage{xspace}
\usepackage{bm}
\usepackage{fancyref}
\usepackage{textcomp}

\usepackage{multirow}
\usepackage{stmaryrd}

\newenvironment{textbmatrix}{	\setlength{\arraycolsep}{2.5pt}%
								\big[\begin{matrix}}{\end{matrix}\big]%
								\raisebox{0.08ex}{\vphantom{M}}}

\def\be{\begin{equation}}
\def\ee{\end{equation}}
\def\een{\nonumber \end{equation}}
\def\mat{\begin{bmatrix}}
\def\emat{\end{bmatrix}}
\def\btm{\begin{textbmatrix}}
\def\etm{\end{textbmatrix}}

\def\ba#1\ea{\begin{align}#1\end{align}}
\def\bas#1\eas{\begin{align*}#1\end{align*}}
\def\bs#1\es{\begin{split}#1\end{split}} 
\def\bg#1\eg{\begin{gather}#1\end{gather}}
\def\bml#1\eml{\begin{multline}#1\end{multline}}
\def\bi#1\ei{\begin{itemize}#1\end{itemize}}

\newcommand{\subtext}[1]{\text{\fontfamily{cmr}\fontshape{n}\fontseries{m}\selectfont{}#1}}
\newcommand{\lefto}{\mathopen{}\left}

\DeclareMathOperator{\Exop}{\opE}			
\DeclareMathOperator{\Varop}{\mathrm{Var}}
\DeclareMathOperator{\Covop}{\mathrm{Cov}}
\DeclareMathOperator{\rot}{\nabla\times}		

\newcommand{\Ex}[2]{\ensuremath{\Exop_{#1}\lefto[#2\right]}} 	
\newcommand{\abs}[1]{\lefto\lvert#1\right\rvert}		

\newcommand{\vecnorm}[1]{\lefto\lVert#1\right\rVert}		

\safemath{\dirac}{\delta}					
\safemath{\krond}{\dirac}					

\safemath{\upto}{\uparrow}
\safemath{\downto}{\downarrow}
\safemath{\iu}{j}							
\safemath{\ev}{\lambda}						
\safemath{\hilseqspace}{l^{2}}				
\newcommand{\banachfunspace}[1]{\setL^{#1}}	
\safemath{\hilfunspace}{\banachfunspace{2}}	

\safemath{\SNR}{\textsf{SNR}} 				
\safemath{\PAR}{\textsf{PAR}} 				
\safemath{\No}{N_0}							
\safemath{\Es}{E_s}							
\safemath{\Eb}{E_b}							
\safemath{\EbNo}{\frac{\Eb}{\No}}
\safemath{\EsNo}{\frac{\Es}{\No}}

\DeclareMathOperator{\CHop}{\ensuremath{\opH}} 
\safemath{\tvir}{\rndh_{\CHop}}				
\safemath{\tvtf}{\rndl_{\CHop}}				
\safemath{\spf}{\rnds_{\CHop}}				
\safemath{\bff}{H_{\CHop}}					

\safemath{\ircf}{r_{h}}						
\safemath{\tftvcf}{r_{s}}					
\safemath{\tfcf}{r_{l}}						
\safemath{\bfcf}{r_{H}}						

\safemath{\tcorr}{c_h}						
\safemath{\scf}{c_{s}}						
\safemath{\tfcorr}{c_{l}}					
\safemath{\fcorr}{c_{H}}						

\safemath{\mi}{I}							
\safemath{\capacity}{C}						

\safemath{\normal}{\mathcal{N}}			
\safemath{\jpg}{\mathcal{CN}}			
\safemath{\mchain}{\leftrightarrow}		
\newcommand{\AIC}{\ensuremath{\text{AIC}}} 

\safemath{\dB}{\,\mathrm{dB}}
\safemath{\dBm}{\,\mathrm{dBm}}
\safemath{\Hz}{\,\mathrm{Hz}}
\safemath{\kHz}{\,\mathrm{kHz}}
\safemath{\MHz}{\,\mathrm{MHz}}
\safemath{\GHz}{\,\mathrm{GHz}}
\safemath{\s}{\,\mathrm{s}}
\safemath{\ms}{\,\mathrm{ms}}
\safemath{\mus}{\,\mathrm{\text{\textmu}s}}
\safemath{\ns}{\,\mathrm{ns}}
\safemath{\ps}{\,\mathrm{ps}}
\safemath{\meter}{\,\mathrm{m}}
\safemath{\mm}{\,\mathrm{mm}}
\safemath{\cm}{\,\mathrm{cm}}
\safemath{\m}{\,\mathrm{m}}
\safemath{\W}{\,\mathrm{W}}
\safemath{\mW}{\, \mathrm{mW}}
\safemath{\J}{\,\mathrm{J}}
\safemath{\K}{\,\mathrm{K}}
\safemath{\bit}{\,\mathrm{bit}}
\safemath{\nat}{\,\mathrm{nat}}

\safemath{\define}{\triangleq}			

\safemath{\equivalent}{\sim}
\safemath{\distas}{\sim}					
\safemath{\sdiff}{\Delta}				

\safemath{\reals}{\mathbb{R}}
\safemath{\positivereals}{\reals_{+}}
\safemath{\integers}{\mathbb{Z}}
\safemath{\posint}{\integers_{+}}
\safemath{\naturals}{\mathbb{N}}
\safemath{\posnaturals}{\naturals_{+}}
\safemath{\complexset}{\mathbb{C}}
\safemath{\rationals}{\mathbb{Q}}

\newcommand*{\fancyrefapplabelprefix}{app}		
\newcommand*{\fancyrefthmlabelprefix}{thm}		
\newcommand*{\fancyreflemlabelprefix}{lem}		
\newcommand*{\fancyrefcorlabelprefix}{cor}		
\newcommand*{\fancyrefdeflabelprefix}{def}		
\newcommand*{\fancyrefproplabelprefix}{prop}	
\newcommand*{\fancyrefobslabelprefix}{obs}		
\newcommand*{\fancyrefalglabelprefix}{alg}		
\newcommand*{\fancyrefasmlabelprefix}{asm}	    
\newcommand*{\fancyreftbllabelprefix}{tab}	 

\frefformat{vario}{\fancyrefseclabelprefix}{Section~#1}
\frefformat{vario}{\fancyrefthmlabelprefix}{Theorem~#1}
\frefformat{vario}{\fancyreflemlabelprefix}{Lemma~#1}
\frefformat{vario}{\fancyrefcorlabelprefix}{Corollary~#1}
\frefformat{vario}{\fancyrefdeflabelprefix}{Definition~#1}
\frefformat{vario}{\fancyrefobslabelprefix}{Observation~#1}
\frefformat{vario}{\fancyrefasmlabelprefix}{Assumption~#1}
\frefformat{vario}{\fancyreffiglabelprefix}{Fig.~#1}
\frefformat{vario}{\fancyrefapplabelprefix}{Appendix~#1} 
\frefformat{vario}{\fancyrefproplabelprefix}{Proposition~#1}
\frefformat{vario}{\fancyrefalglabelprefix}{Algorithm~#1}
\frefformat{vario}{\fancyreftbllabelprefix}{Table~#1}
\frefformat{vario}{\fancyrefeqlabelprefix}{(#1)}

%% file: DAB_J_Resub_v1.bbl
\begin{thebibliography}{10}
\providecommand{\url}[1]{#1}
\csname url@samestyle\endcsname
\providecommand{\newblock}{\relax}
\providecommand{\bibinfo}[2]{#2}
\providecommand{\BIBentrySTDinterwordspacing}{\spaceskip=0pt\relax}
\providecommand{\BIBentryALTinterwordstretchfactor}{4}
\providecommand{\BIBentryALTinterwordspacing}{\spaceskip=\fontdimen2\font plus
\BIBentryALTinterwordstretchfactor\fontdimen3\font minus
  \fontdimen4\font\relax}
\providecommand{\BIBforeignlanguage}[2]{{%
\expandafter\ifx\csname l@#1\endcsname\relax
\typeout{** WARNING: IEEEtran.bst: No hyphenation pattern has been}%
\typeout{** loaded for the language `#1'. Using the pattern for}%
\typeout{** the default language instead.}%
\else
\language=\csname l@#1\endcsname
\fi
#2}}
\providecommand{\BIBdecl}{\relax}
\BIBdecl

\bibitem{aghdam}
S.~Rezaei~Aghdam, S.~Jacobsson, and T.~Eriksson, ``Distortion-aware linear
  precoding for millimeter-wave multiuser {MISO} downlink,'' in \emph{{IEEE}
  Int. Conf. Comm. (ICC) workshops}, Shanghai, China, May 2019.

\bibitem{bjornson_book}
E.~Bj\"{o}rnson, J.~Hoydis, and L.~Sanguinetti, ``Massive {MIMO} networks:
  Spectral, energy, and hardware efficiency,'' \emph{Found. Trends Signal
  Process.}, vol.~11, no. 3–4, pp. 154--655, Nov. 2017.

\bibitem{gustavsson14a}
U.~Gustavsson, C.~Sanch{\'e}z-Perez, T.~Eriksson, F.~Athley, G.~Durisi,
  P.~Landin, K.~Hausmair, C.~Fager, and L.~Svensson, ``On the impact of
  hardware impairments on massive {MIMO},'' in \emph{Proc. IEEE Global Commun.
  Conf. (GLOBECOM)}, Austin, TX, USA, Dec. 2014, pp. 294--300.

\bibitem{bjornson14a}
E.~Bj{\"o}rnson, J.~Hoydis, M.~Kountouris, and M.~Debbah, ``Massive {MIMO}
  systems with non-ideal hardware: Energy efficiency, estimation, and capacity
  limits,'' \emph{{IEEE} Trans. Inf. Theory}, vol.~11, no.~60, pp. 7112--7139,
  Nov. 2014.

\bibitem{Papazaf17}
A.~{Papazafeiropoulos}, B.~{Clerckx}, and T.~{Ratnarajah}, ``Rate-splitting to
  mitigate residual transceiver hardware impairments in massive {MIMO}
  systems,'' \emph{IEEE Trans. on Veh. Technol.}, vol.~66, no.~9, pp.
  8196--8211, Sep. 2017.

\bibitem{abdelghany19}
\BIBentryALTinterwordspacing
M.~Abdelghany, A.~A. Farid, U.~Madhow, and M.~J.~W. Rodwell, ``A design
  framework for all-digital mmwave massive {MIMO} with per-antenna
  nonlinearities,'' Dec. 2019. [Online]. Available:
  \url{https://arxiv.org/abs/1912.11643}
\BIBentrySTDinterwordspacing

\bibitem{qi12a}
J.~Qi and S.~Aissa, ``On the power amplifier nonlinearity in {MIMO} transmit
  beamforming systems,'' \emph{{IEEE} Trans. Commun.}, vol.~60, no.~3, pp.
  876--887, Mar. 2012.

\bibitem{blandino17a}
S.~Blandino, C.~Desset, A.~Bourdoux, L.~Van~der Perre, and S.~Pollin,
  ``Analysis of out-of-band interference from saturated power amplifiers in
  massive {MIMO},'' in \emph{Proc. Eur. Conf. Netw. Commun. (EuCNC)}, Oulu,
  Finland, Jul. 2017.

\bibitem{mollen18b}
C.~Moll{\'e}n, U.~Gustavsson, T.~Eriksson, and E.~G. Larsson, ``Spatial
  characteristics of distortion radiated from antenna arrays with transceiver
  nonlinearities,'' \emph{{IEEE} Trans. Wireless Commun.}, vol.~17, no.~10, pp.
  6663--6679, Oct. 2018.

\bibitem{moghadam18a}
N.~N. Moghadam, G.~Fodor, M.~Bengtsson, and D.~J. Love, ``On the energy
  efficiency of {MIMO} hybrid beamforming for millimeter-wave systems with
  nonlinear power amplifiers,'' \emph{{IEEE} Trans. Wireless Commun.}, vol.~17,
  no.~11, pp. 7208--7221, Nov. 2018.

\bibitem{khanzadi15a}
M.~R. Khanzadi, G.~Durisi, and T.~Eriksson, ``Capacity of {SIMO} and {MISO}
  phase-noise channels with common/separate oscillators,'' \emph{{IEEE} Trans.
  Commun.}, vol.~63, no.~9, pp. 3218--3231, Sep. 2015.

\bibitem{kolomvakis16a}
N.~Kolomvakis, M.~Matthaiou, and M.~Coldrey, ``{IQ} imbalance in multiuser
  systems: Channel estimation and compensation,'' \emph{{IEEE} Trans. Commun.},
  vol.~64, no.~7, pp. 3039--3051, Jul. 2016.

\bibitem{jacobsson17d}
S.~Jacobsson, G.~Durisi, M.~Coldrey, T.~Goldstein, and C.~Studer, ``Quantized
  precoding for massive {MU-MIMO},'' \emph{{IEEE} Trans. Commun.}, vol.~65,
  no.~11, pp. 4670--4684, Nov. 2017.

\bibitem{zhang15d}
X.~Zhang, M.~Matthaiou, M.~Coldrey, and E.~Bj{\"o}rnson, ``Impact of residual
  transmit {RF} impairments on training-based {MIMO} systems,'' \emph{{IEEE}
  Trans. Commun.}, vol.~63, no.~8, pp. 2899--2911, Aug. 2015.

\bibitem{Papazafeiropoulos18}
A.~{Papazafeiropoulos}, S.~K. {Sharma}, T.~{Ratnarajah}, and S.~{Chatzinotas},
  ``Impact of residual additive transceiver hardware impairments on
  rayleigh-product {MIMO} channels with linear receivers: Exact and asymptotic
  analyses,'' \emph{{IEEE} Trans. Commun.}, vol.~66, no.~1, pp. 105--118, Jan.
  2018.

\bibitem{studer10b}
C.~Studer, M.~Wenk, and A.~Burg, ``{MIMO} transmission with residual
  transmit-{RF} impairments,'' in \emph{Proc. Int. ITG Workshop on Smart
  Antennas (WSA)}, Bremen, Germany, Feb. 2010, pp. 189--196.

\bibitem{larsson18a}
E.~G. Larsson and L.~Van~der Perre, ``Out-of-band radiation from antenna arrays
  clarified,'' \emph{{IEEE} Wireless Commun. Lett.}, vol.~7, no.~4, pp.
  610--613, Feb. 2018.

\bibitem{anttila19}
L.~{Anttila}, A.~{Brihuega}, and M.~{Valkama}, ``On antenna array out-of-band
  emissions,'' \emph{IEEE Wireless Commun. Lett.}, vol.~8, no.~6, pp.
  1653--1656, Dec. 2019.

\bibitem{jacobsson18d}
S.~Jacobsson, U.~Gustavsson, G.~Durisi, and C.~Studer, ``Massive
  {MU}-{MIMO}-{OFDM} uplink with hardware impairments: Modeling and analysis,''
  in \emph{Proc. Asilomar Conf. Signals, Syst., Comput.}, Pacific Grove, CA,
  USA, Oct. 2018, pp. 1829--1835.

\bibitem{bjornson19a}
E.~Bj{\"o}rnson, L.~Sanguinetti, and J.~Hoydis, ``Hardware distortion
  correlation has negligible impact on {UL} massive {MIMO} spectral
  efficiency,'' \emph{{IEEE} Trans. Commun.}, vol.~67, no.~2, pp. 1085--1098,
  Feb. 2019.

\bibitem{aghdam19b}
S.~R. Aghdam and T.~Eriksson, ``On the performance of distortion-aware linear
  receivers in uplink massive {MIMO} systems,'' in \emph{Proc. IEEE Int. Symp.
  Wirel. Comm. Syst. (ISWCS)}, Oulu, Finland, Aug. 2019, pp. 208--212.

\bibitem{Demir19}
O.~T. {Demir} and E.~{Bj\"ornson}, ``Channel estimation in massive {MIMO} under
  hardware non-linearities: Bayesian methods versus deep learning,'' \emph{IEEE
  Open J. Commun. Soc.}, vol.~1, pp. 109--124, Jan. 2020.

\bibitem{sheikhi2021}
A.~Sheikhi, F.~Rusek, and O.~Edfors, ``Massive mimo with per-antenna digital
  predistortion size optimization: Does it help?'' in \emph{Proc. IEEE Int.
  Conf. Commun. (ICC)}, Montreal, QC, Canada, Jun. 2021, pp. 1--6.

\bibitem{brihuega_asil}
A.~Brihuega, L.~Anttila, M.~Abdelaziz, and M.~Valkama, ``Digital predistortion
  in large-array digital beamforming transmitters,'' in \emph{Proc. Asilomar
  Conf. Signals, Syst., Comput.}, Pacific Grove, CA, USA, Feb. 2018, pp.
  611--618.

\bibitem{abdelaziz17}
M.~Abdelaziz, L.~Anttila, and M.~Valkama, ``Reduced-complexity digital
  predistortion for massive {MIMO},'' in \emph{Proc. IEEE Int. Conf. Acoust.,
  Speech, Signal Process. (ICASSP)}, New Orleans, LA, USA, Jun. 2017, pp.
  6478--6482.

\bibitem{bussgang52a}
J.~J. Bussgang, ``Crosscorrelation functions of amplitude-distorted {Gaussian}
  signals,'' Res. Lab. Elec., Cambridge, MA, USA, Tech. Rep. 216, Mar. 1952.

\bibitem{schenk2008rf}
T.~Schenk, \emph{RF imperfections in high-rate wireless systems: {I}mpact and
  digital compensation}.\hskip 1em plus 0.5em minus 0.4em\relax Dordrecht, The
  Netherlands: Springer-Verlag, 2008.

\bibitem{Persson13}
D.~{Persson}, T.~{Eriksson}, and E.~G. {Larsson}, ``Amplifier-aware
  multiple-input multiple-output power allocation,'' \emph{IEEE Commun. Lett.},
  vol.~17, no.~6, pp. 1112--1115, May 2013.

\bibitem{grebennikov2005rf}
A.~Grebennikov, \emph{RF and microwave power amplifier design}.\hskip 1em plus
  0.5em minus 0.4em\relax McGraw-Hill New York, 2005.

\bibitem{reed62a}
I.~S. Reed, ``On a moment theorem for complex {Gaussian} processes,''
  \emph{{IEEE} Trans. Inf. Theory}, vol.~8, no.~3, pp. 194--195, Apr. 1962.

\bibitem{lapidoth96b}
A.~Lapidoth, ``Nearest neighbor decoding for additive non-{Gaussian} noise
  channels,'' \emph{{IEEE} Trans. Inf. Theory}, vol.~42, no.~5, pp. 1520--1529,
  Sep. 1996.

\bibitem{arnold06a}
D.~M. Arnold, H.-A. Loeliger, P.~O. Vontobel, A.~Kavcic, and W.~Zeng,
  ``Simulation-based computation of information rates for channels with
  memory,'' \emph{{IEEE} Trans. Inf. Theory}, vol.~52, no.~8, pp. 3498--3508,
  Aug. 2006.

\bibitem{lapidoth96a}
A.~Lapidoth, ``Mismatched decoding and the multiple access channel,''
  \emph{{IEEE} Trans. Inf. Theory}, vol.~42, no.~5, pp. 1439--1452, Sep. 1996.

\bibitem{palomar05gradient}
D.~P. Palomar and S.~Verd{\'u}, ``Gradient of mutual information in linear
  vector gaussian channels,'' \emph{{IEEE} Trans. Inf. Theory}, vol.~52, no.~1,
  pp. 141--154, Aug. 2005.

\bibitem{bertsekas1997nonlinear}
D.~P. Bertsekas, ``Nonlinear programming,'' \emph{J. Oper. Res. Soc.}, vol.~48,
  no.~3, pp. 334--334, 1997.

\bibitem{feng13survey}
D.~{Feng}, C.~{Jiang}, G.~{Lim}, L.~J. {Cimini}, G.~{Feng}, and G.~Y. {Li}, ``A
  survey of energy-efficient wireless communications,'' \emph{IEEE Commun.
  Surv. Tut.}, vol.~15, no.~1, pp. 167--178, Feb. 2013.

\bibitem{alkhateeb15a}
A.~Alkhateeb, G.~Leus, and R.~W. {Heath Jr.}, ``Limited feedback hybrid
  precoding for multi-user millimeter wave systems,'' \emph{{IEEE} Trans.
  Wireless Commun.}, vol.~14, no.~11, pp. 6481--6494, Nov. 2015.

\bibitem{akdeniz14a}
M.~R. Akdeniz, Y.~Liu, M.~K. Samimi, S.~Sun, S.~Rangan, T.~S. Rappaport, and
  E.~Erkip, ``Millimeter wave channel modeling and cellular capacity
  evaluation,'' \emph{{IEEE} J. Sel. Areas Commun.}, vol.~32, no.~6, pp.
  1164--1179, Jun. 2014.

\bibitem{jee2020precoding}
J.~Jee, G.~Kwon, and H.~Park, ``Precoding design and power control for {SINR}
  maximization of {MISO} system with nonlinear power amplifiers,'' \emph{IEEE
  Trans. on Veh. Technol.}, vol.~69, no.~11, pp. 14\,019--14\,024, Sep. 2020.

\end{thebibliography}
